\documentclass[prd,preprint,tightenlines,floatfix,showpacs,preprintnumbers,nofootinbib,eqsecnum,superscriptaddress]{revtex4-1}

\usepackage{color}      
\usepackage[dvips]{graphicx}    
\usepackage{amsfonts}      
\usepackage{amsmath}       
\usepackage{amssymb}       
\usepackage{psfrag}

\newcommand*{\sub}[1]{_{\mathrm{#1}}}

\newcommand{\sinb}{\sin\beta}
\newcommand{\cosb}{\cos\beta}

\renewcommand{\sb}{s_\beta}
\newcommand{\cb}{c_\beta}

\newcommand{\muF}{\mu_{\mathrm{F}}}
\newcommand{\muR}{\mu_{\mathrm{R}}}
\newcommand{\alphas}{\alpha_{\mathrm{s}}}
\newcommand{\GF}{G_{\mathrm{F}}}
\newcommand{\shat}{\hat{s}}
\newcommand{\that}{\hat{t}}
\newcommand{\uhat}{\hat{u}}
\newcommand{\DRbar}{\overline{\mathrm{DR}}}
\newcommand{\MSbar}{\overline{\mathrm{MS}}}
\newcommand{\mueff}{\mu_{\mathrm{eff}}}

\newcommand{\sigbbhw}{\hat{\sigma}_{b\bar{b}\to H^\pm W^\mp}}
\newcommand{\siggghw}{\hat{\sigma}_{gg\to H^\pm W^\mp}}
\newcommand{\sigpphw}{\sigma_{pp\to H^\pm W^\mp}}

\newcommand{\GeV}{\;\mathrm{GeV}}
\newcommand{\TeV}{\;\mathrm{TeV}}

\newcommand{\NMT}{{\sc NMSSMTools}}

\newcommand{\FA}{{\sc FeynArts}}
\newcommand{\FC}{{\sc FormCalc}}
\newcommand{\LT}{{\sc LoopTools}}
\newcommand{\HB}{{\sc HiggsBounds}}

\newcommand{\MHp}{{m_{H^\pm}}}
\newcommand{\MHpsq}{{m^2_{H^\pm}}}
\newcommand{\hp}{{H^\pm}}
\newcommand{\HW}{{H^\pm W^\mp}}
\newcommand{\f}{\frac}
\newcommand{\nn}{\nonumber}
\newcommand{\si}{\mathrm{i}}

\newcommand{\reffig}[1]{{Fig.~\ref{fig:#1}}}

\begin{document}

\begin{flushright}
DESY 11-247 \\
 LU TP 11-47\\
 December 2011
\end{flushright}
\vspace{1cm}

\title{Enhancement of associated $H^\pm W^\mp$ production in the NMSSM}


\author{R.~Enberg}
 \email{Rikard.Enberg@physics.uu.se}
 \affiliation{
 Department of Physics and Astronomy,
 Uppsala University, Box 516, SE--751 20 Uppsala, Sweden}

\author{R.~Pasechnik}
 \email{Roman.Pasechnik@thep.lu.se}
 \affiliation{
 Department of Astronomy and Theoretical Physics,
 Lund University, SE--223 62 Lund, Sweden}

\author{O.~St{\aa}l}
\email{oscar.stal@desy.de}
\affiliation{
 Deutsches Elektronen-Synchrotron DESY,
 Notkestra{\ss}e 85, D--22607 Hamburg, Germany}

\begin{abstract}
We study the associated production of a charged Higgs and a $W$ boson in high-energy
$pp$ collisions at the Large Hadron Collider. This is an interesting process for charged Higgs discovery, or exclusion, since
the production cross section could depend strongly on the model, offering potential discriminating power between supersymmetric extensions of the Standard Model with minimal or extended Higgs sectors.
We compute the cross section for this process in the next-to-minimal supersymmetric Standard Model (NMSSM), at the tree level for quark-quark scattering and at one-loop level for gluon-gluon scattering. The most important corrections beyond leading order are taken into account using an improved Born approximation. We find that the $pp\to H^\pm W^\mp$ cross section can be resonantly enhanced by up to an order of
magnitude over its MSSM value (both for $\sqrt{s}=7\TeV$ and $14\TeV$) through the
contributions of heavy, neutral, singlet-dominated Higgs bosons appearing
in the $s$-channel. Since such Higgs mass configurations are normally not
possible in the MSSM, the observation of associated $H^\pm W^\mp$ production  at the LHC  could provide a striking, although indirect, signature of a more complicated Higgs sector.
\end{abstract}


\maketitle

\section{Introduction}

There are good phenomenological and theoretical reasons to consider supersymmetric (SUSY) extensions of the Standard Model, including amelioration of the hierarchy problem or the fine-tuning of the Higgs mass, unification of the gauge couplings, and the existence of a dark matter WIMP candidate. The most common SUSY model is the minimal supersymmetric extension of the Standard Model (MSSM) \cite{Nilles:1983ge,Haber:1984rc}, which exhibits all of the above benefits. These benefits come at the price of introducing a mass term (called the $\mu$-term) for the two Higgs doublets that are required in the MSSM. This leads to another fine-tuning problem, namely why the mass parameter $\mu$ of the Higgs doublets should be at the electroweak mass scale, as is required by phenomenology. This is known as the $\mu$-problem. Furthermore, the mass of the lightest Higgs boson cannot be less than $\sim 115$~GeV, while in the MSSM at the tree-level it has an upper limit equal to the $Z$ boson mass. To fulfill the experimental constraints, large corrections to the Higgs mass from top- and stop-loops are needed, which require a rather large stop mass leading to an additional amount of fine-tuning in the model. If we take fine-tuning arguments seriously, we may therefore consider whether there are alternatives to the MSSM (see e.g.\ \cite{Barbieri:2006bg,HallPinnerRuderman}).

One such alternative is the next-to-minimal supersymmetric Standard Model (NMSSM), which has recently been reviewed in Refs.\ \cite{Ellwanger:2009dp,Maniatis:2009re}. In the NMSSM, there is in addition to the two Higgs doublets of the MSSM a singlet Higgs field, which is the scalar component of a chiral singlet superfield added to the MSSM superpotential. The reason for introducing this additional scalar is that the $\mu$-term is now dynamically generated, so that the fine-tuned parameter $\mu$ is no longer needed.
This scalar mixes with the other scalars from the two doublets, leading to a Higgs sector with seven Higgs bosons, compared to the five present in the MSSM. The fermion component of the singlet superfield, the singlino, additionally mixes with the neutralinos, providing interesting possibilities for dark matter that can be, e.g., singlino or singlino--Higgsino-dominated \cite{Ellwanger:2009dp}, or very light \cite{Draper:2010ew,Vasquez:2010ru}.

There are two extra neutral Higgs bosons in the NMSSM compared to the MSSM; one CP-even and one CP-odd. The tree-level mass relations for the Higgs bosons are then also modified, and it is possible for one or more Higgs bosons to be quite light. In
particular the lightest CP-odd Higgs, $A_1$, can be significantly lighter than in the MSSM---viable scenarios with $m_{A_1}<2m_b$ exist~\cite{Dermisek:2005ar}. The charged Higgs boson can also be rather light~\cite{Akeroyd:2007yj,Mahmoudi:2010xp}, albeit not as light as the $A_1$.

Because of the different Higgs phenomenology due to the modified mass relations and the additional particles, the parameters of the model are not as constrained as in the MSSM, and new decay channels and production mechanisms may become important at the LHC. For example, the charged Higgs may decay as $H^+\to W^+ A_1$, and the $A_1$ in turn may decay dominantly as $A_1\to b\bar b, \tau^+\tau^-$. Such differences are important to take into account in searches at LHC, so that no possibilities are missed.

In this paper, our focus is on production of the charged Higgs boson $H^\pm$ in association with a $W^\mp$ boson. This is not the main production channel usually considered for $H^\pm$ and, to the best of our knowledge, it has not been studied in the context of the NMSSM. It was however pointed out in Ref.~\cite{Enberg:2011qh} that a related process, the associated central exclusive production of $\HW$, may be useful in the NMSSM.

The usual production mechanisms for $\hp$ are top quark decays $t\to H^+ b$ for light charged Higgs (where the charged Higgs boson mass $\MHp<m_t$), and production with a top, $bg\to H^-t$ or $gg\to H^-t\bar b$, for heavy charged Higgs ($\MHp>m_t$). The associated production mechanism may, however, be important to secure additional information about the Higgs sector if the charged Higgs is first observed in one of the above-mentioned processes. An advantage of the $\HW$  process is that a leptonically decaying $W$ may be used as an experimental requirement. The cross section for associated production is as we shall see rather model-dependent, and the observation of this process may therefore provide constraints on the model parameters. For example, in the NMSSM, smaller $\tan\beta$ is allowed, leading to a possible enhancement of the cross section. 
A potentially more important difference between the MSSM results and the NMSSM comes from the resonant $s$-channel exchanges of additional singlet-dominated Higgs bosons. Due to the very restricted mass relations of the MSSM Higgs sector, these contributions cannot be resonant in the MSSM, while in the NMSSM they can. The resonant enhancement of the parton-level cross section will also enhance the hadron-level cross section in some range of charged Higgs boson masses and this enhancement could be potentially visible at the LHC. It could be important for discerning differences between MSSM and NMSSM, and for setting limits on the parameter space.

The $H^\pm W^\mp$ production channel was first considered in \cite{Dicus:1989vf}, where the cross sections were calculated in the approximation that $m_b=0$. This study did not include any contribution from squark loops. The cross sections were later calculated in full generality for two-Higgs-doublet models (2HDM) and the MSSM in \cite{BarrientosBendezu:1998gd,BarrientosBendezu:1999vd,BarrientosBendezu:2000tu,Brein:2000cv}, where in particular \cite{BarrientosBendezu:2000tu,Brein:2000cv} included the squark loop contributions. In \cite{Asakawa:2005nx}, it was further shown that there may be a substantial enhancement of the cross section compared to the MSSM in a general 2HDM.
Our paper extends these studies to the NMSSM, and we are going to investigate in particular the differences between the MSSM and the NMSSM for this process.

Next-to-leading order (NLO) corrections to the $b\bar b\to \HW$ subprocess, including QCD, SUSY QCD, and electroweak contributions, are known for the MSSM~\cite{Yang:2000yt,Hollik:2001hy,Zhao:2005mu,Gao:2007wz,Rauch:2008fy,Dao:2010nu}. There have also been phenomenological studies of this process for the LHC at 14~TeV \cite{Moretti:1998xq,Eriksson:2006yt,Gao:2007wz,Hashemi:2010ce,Bao:2011sy}. While we do discuss the LHC aspects of the $H^\pm W^\mp$ process in the NMSSM below, the inclusion of NLO corrections and a detailed study of LHC signatures are beyond the scope of this paper and will be left for future studies.

Finally, we restrict ourselves to the case of charged Higgs bosons heavier than the top quark, which means that the $t\to b H^+$ production mechanism is not effective. The reason is twofold:\ first, this decay is the same in the MSSM and the NMSSM, but we are interested in differences between the two models. Second, the experimental constraints on this decay channel are already quite strict~\cite{ATLAS-CONF-2011-138,ATLAS-CONF-2011-151,CMS-PAS-HIG-11-008}, ruling out $\mathrm{BR}(t\to bH^+)>5-10\%$  (in the MSSM).

\section{The NMSSM Higgs sector}

The NMSSM is defined by removing the $\mu$-term from the MSSM
superpotential and adding a singlet chiral superfield $\hat S$,
which only couples to the Higgs doublets. Assuming scale invariance,
the general form of the superpotential is
\begin{equation}
 W\sub{NMSSM} = W\sub{MSSM} + \lambda \hat S \hat H_u \hat H_d + \frac{\kappa}{3} \hat S^3.
\end{equation}
In the above superpotential, the (unchanged) Yukawa terms are
contained in $W\sub{MSSM}$. The scalar potential of the NMSSM is
obtained from the $F$- and $D$-terms plus the soft SUSY-breaking
terms for the Higgs sector,
\begin{equation}
 V\sub{soft} = m_{H_u}^2 \left|H_u\right|^2 + m_{H_d}^2 \left|H_d\right|^2 + m_S^2 \left|S\right|^2
+ \left[ \lambda A_\lambda S H_u H_d +\frac{1}{3}\kappa A_\kappa S^3 + \text{h.c.}\right],
\end{equation}
where the dimensionless couplings $\lambda$ and $\kappa$, the soft
SUSY-breaking parameters $A_\lambda$ and $A_\kappa$ with dimension
of mass, and the singlet mass $m_S$ are new parameters compared to
the MSSM. As usual, $m_S$ is fixed by the minimization of the
potential. Requiring that $S$ acquires a vacuum expectation value
(vev), $s=\left\langle S\right\rangle$, yields an additional new
parameter of the model, and gives rise to an effective
$\mu$-parameter $\mu\sub{eff}=\lambda s$. Together with the ratio of
vevs of the two Higgs doublets, $\tan\beta=v_u/v_d$, where
$v_u^2+v_d^2=v^2=(174\text{ GeV})^2$, we have six free
parameters of the Higgs sector of the NMSSM: $\lambda, \kappa,
A_\lambda, A_\kappa, s$, and $\tan\beta$.

As $S$ is a complex field, there will be two additional physical
Higgs bosons in the NMSSM compared to the MSSM. For a CP-conserving
theory (as is assumed here) we have three CP-even neutral states
$H_1, H_2, H_3$ and two CP-odd neutral states $A_1$ and $A_2$, where
we take the states to be ordered in mass with $H_1$ and $A_1$ the
lightest states.

The mass of the charged Higgs boson is at tree-level
\begin{equation}
m_{H^\pm}^2 = \frac{2\mu\sub{eff}}{\sin 2\beta}(A_\lambda + \kappa
s) + m_W^2 -\lambda^2 v^2, \label{eq:mhp2}
\end{equation}
which can be compared with the MSSM expression $m_{H^\pm}^2 = m_A^2
+ m_W^2$. It therefore simplifies our expressions to define an
effective ``doublet mass'' in the NMSSM as
\begin{equation}
 m_A^2 = \frac{2\mu\sub{eff}}{\sin 2\beta}(A_\lambda + \kappa s) = \frac{\lambda s}{s_\beta c_\beta}(A_\lambda + \kappa s)
       = \frac{\mu\sub{eff}}{s_\beta c_\beta}(A_\lambda  + \frac{\kappa}{\lambda} \mu\sub{eff}),
\label{eq:ma2}
\end{equation}
where we defined $s_\beta=\sinb$ and $c_\beta=\cosb$. Thus, for
fixed $\mu\sub{eff}$, the squared doublet mass and the charged Higgs mass
both depend linearly on $A_\lambda$ and on $\kappa/\lambda$. The
parameter $A_\lambda$ may be everywhere swapped for $m_A$ as a
parameter of the Higgs sector. The mass relation \eqref{eq:mhp2} can
then be written as
\begin{equation}
 m_{H^\pm}^2 = m_A^2 + m_W^2 - \lambda^2 v^2.
 \label{eq:mhc2}
\end{equation}
This exhibits one important difference between the NMSSM and the
MSSM: in the MSSM, the charged Higgs and the CP-odd Higgs are almost
degenerate in mass as soon as they are heavier than $\sim 200$~GeV.
In the NMSSM there is no such strong correlation; partly because of
the additional contribution $-\lambda^2 v^2$ which lowers
$m_{H^\pm}^2$, but mainly because there is usually no physical state
with mass $m_A$. The two CP-odd states in the NMSSM arise as
mixtures of the CP-odd state of the MSSM, $A=\text{Im} (H_u\cb + H_d
\sb)$, and of the imaginary part of the scalar $S$. The effective
mass $m_A$ thus only corresponds to a physical mass if the mixing
between the two pseudoscalar bosons vanishes. This mixing is
obtained from the mass matrix for $A_1,\,A_2$,
\begin{equation}
 \mathcal{M}_P^2 =
\begin{pmatrix}
 m_A^2 & \dfrac{v}{s} \left(m_A^2 \sb\cb - 3\lambda\kappa s^2\right) \\
\dfrac{v}{s} \left(m_A^2 \sb\cb - 3\lambda\kappa s^2\right) &
\dfrac{v^2}{s^2}\sb\cb \left(m_A^2 \sb\cb + 3\lambda\kappa
s^2\right) -3\kappa A_\kappa s
\end{pmatrix}\,.
\end{equation}
In terms of the weak basis eigenstates $A_j^{\rm
weak}=(\text{Im}\,H_d,\,\text{Im}\,H_u,\,\text{Im}\,S)$, the
physical CP-odd eigenstates $A_i^{\rm mass}=(A_1,\,A_2)$ (ordered in
increasing mass) are given by $A_i^{\rm mass}=P_{ij}A_j^{\rm weak}$
with the $2\times3$ mixing matrix $P_{ij}$, or explicitly
$$\binom{A_1}{A_2}=
\begin{pmatrix}
 P_{11} & P_{12} & P_{13} \\
 P_{21} & P_{22} & P_{23}
\end{pmatrix}
\begin{pmatrix}
\text{Im}\,H_d \\
\text{Im}\,H_u \\
\text{Im}\,S
\end{pmatrix}\,.
$$
It has been argued, for example by Dermisek and
Gunion~\cite{Dermisek:2005ar,Dermisek:2008uu,Dermisek:2009fd,Dermisek:2010mg},
that $A_1$ may be much lighter than the other Higgs bosons, and can
even be as light as a few GeV, and still be allowed by EWPT and
collider constraints.

The masses of the CP-even Higgs states require a three-dimensional
mixing matrix $S_{ij}$ rotating the weak basis $H_j^{\rm
weak}=(\text{Re}\,H_d,\,\text{Re}\,H_u,\,\text{Re}\,S)$ to the
physical one,
$$H_i^{\rm mass}=S_{ij}H_j^{weak}\,,$$ such that the physical mass
eigenstates $H_i^{\rm mass}$ are ordered in increasing mass. The
corresponding expressions for masses and mixings are omitted here;
they may be found in e.g.~\cite{Ellwanger:2009dp}. It can be
shown that the tree-level mass of the lightest Higgs boson $H_1$ is
no longer limited by $m_Z$ as in the MSSM, but instead by
\begin{equation}
 m_{H_1}^2 \le m_Z^2\cos^2 2\beta + \lambda^2 v^2 \sin^2 2\beta.
\end{equation}
To summarize, in the NMSSM (at tree-level), the lightest CP-even
Higgs is allowed to be somewhat heavier, and the charged Higgs
somewhat lighter, than in the MSSM, while the lightest CP-even Higgs
may be much lighter than in the MSSM.

In the above discussion we have considered only the tree-level
masses, but just as in the MSSM there can be considerable
corrections to these masses at higher orders \cite{Ellwanger:1993hn,Elliott:1993ex,Elliott:1993uc,Elliott:1993bs,Pandita:1993hx,Ellwanger:2005fh,Staub:2010ty,Ender:2011qh}. To
take these into account to the best precision available
\cite{Degrassi:2009yq}, we use the code \NMT{} (version 2.3.5)
\cite{Ellwanger:2004xm,Ellwanger:2005dv} for our numerical
evaluation of the Higgs mass spectrum. We also use this code in the
following to calculate the mixing, all coupling strengths, and the
Higgs decay widths from the given input parameters.

Beyond leading order, the Higgs spectrum depends on all the
parameters listed above, as well as on the soft SUSY-breaking
parameters of other sectors; the most important corrections
typically come from stop mixing. The standard way to cope with this
situation in the MSSM is to consider a benchmark scenario (such as those
defined in \cite{Carena:1999xa,Carena:2002qg}) to fix the higher
order corrections from other SUSY sectors, and then vary
independently the parameters in the Higgs sector. We shall use the
same approach here, extending in a straightforward manner the MSSM
benchmark scenario to the NMSSM. We therefore use $\mu\sub{eff}$ as
an input, which together with the NMSSM coupling $\lambda$
determines the value of the singlet vev $s$. Since we are mainly
interested in comparing the NMSSM to the MSSM, rather than the MSSM
to itself, we use the same benchmark (inspired by maximal mixing)
throughout this work:
\begin{equation*}
\begin{aligned}
M_{\mathrm{SUSY}}&=1\TeV,\quad X^{\mathrm{\overline{DR}}}_t\equiv A_t-\mueff\cot\beta=\sqrt{6}M_{\mathrm{SUSY}},\quad A_b=A_\tau=A_t\\\mueff&=250\GeV,\quad
M_1=100\GeV,\quad M_2=200\GeV,\quad M_3=1\TeV
\end{aligned}
\end{equation*}
For the remaining (free) input parameters of the NMSSM Higgs sector,
we shall take $A_\lambda$, $\tan\beta$---which are equivalent to
the two parameters $\MHp$, $\tan\beta$ in the MSSM through
Eqs.~(\ref{eq:ma2}), (\ref{eq:mhc2})---and in addition $\lambda$,
$\kappa$, and $A_{\kappa}$, which are specific to NMSSM. For any
choice of NMSSM parameters, the corresponding MSSM limit can be
obtained by taking $\lambda\to 0$, $\kappa\to 0$, while keeping the
ratio $\kappa/\lambda$ and all dimensionful parameters fixed.
\newpage

\section{Associated $H^\pm W^\mp$ production}
\label{sect:theory}
\begin{figure}[t!]
 \centerline{\includegraphics[width=0.5\textwidth]{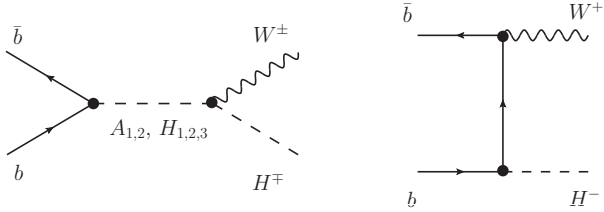}}
 \caption{Leading-order diagrams for the quark-initiated hard subprocess $b\bar b\to
H^{\pm}W^{\mp}$.} \label{fig:bbHW-diag}
\end{figure}
\begin{figure}[th!]
\psfrag{H}[Bc][Bc]{$H^\pm\,\,$}
\psfrag{W}[Bc][Bc]{$W^\mp\,\,$}
\psfrag{g}[Bc][Bc]{$g$}
\centerline{\includegraphics[width=0.8\textwidth]{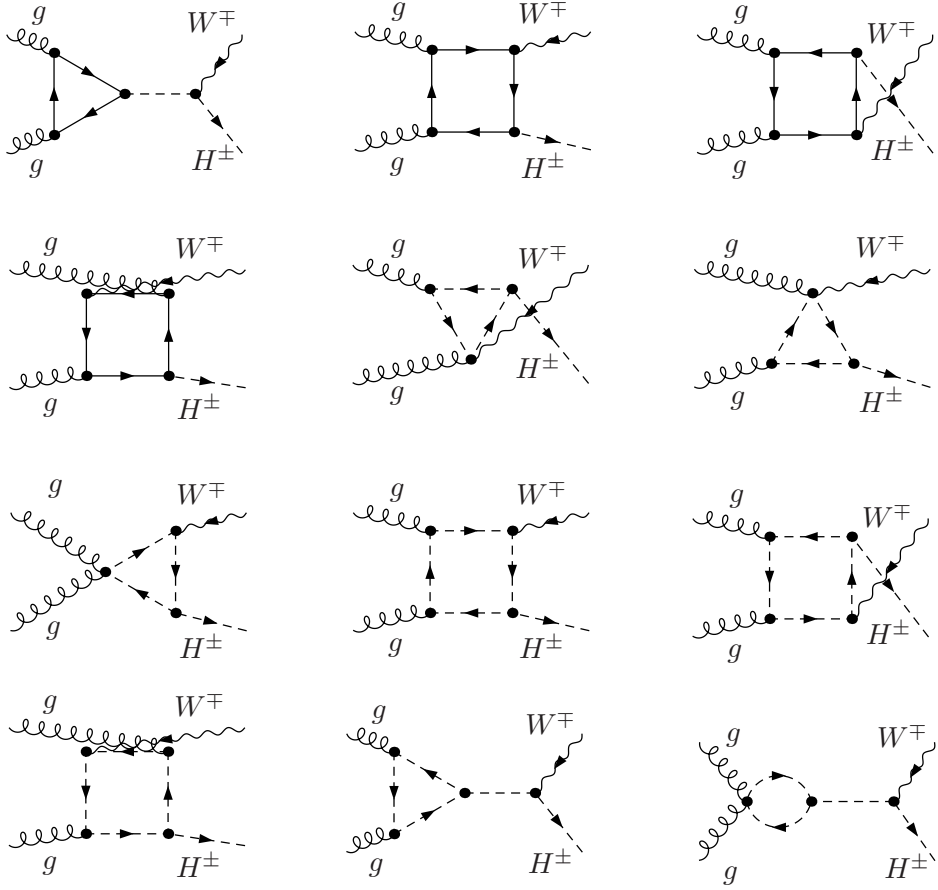}}
\caption{Example leading-order diagrams  contributing to the gluon-initiated
hard subprocess $gg\to H^{\pm}W^{\mp}$. Dashed lines with arrows represent squarks, while internal dashed lines without arrows represent neutral Higgs bosons ($H_{1,2,3},A_{1,2}$). Only the diagrams with internal $s$-channel Higgs propagators will give rise to the resonant enhancement we are discussing in this paper.} \label{fig:ggHW-diag}
\end{figure}

Associated $\HW$ production has two contributing subprocesses at leading order, quark-antiquark ($q\bar{q}$) and gluon-gluon ($gg$) scattering. The leading order contributions correspond to the tree-level for $q\bar{q}$, and one loop for $gg$ fusion. Representative diagrams for these subprocesses are depicted in Figs.~\ref{fig:bbHW-diag} and \ref{fig:ggHW-diag}, respectively. Working in a five-flavor scheme (5FS) with an effective parton distribution for the $b$ quark, the $q\bar{q}$ process is completely dominated by the $b\bar b$ contribution. 
Although the cross section for $gg$ fusion is formally
suppressed by two powers of the QCD coupling $\alphas$ relative to
$b\bar{b}$ annihilation, it may yield a comparable contribution at
LHC energies due to the large gluon density at small $x$ and needs to be taken into account.
In the 5FS, there are additional contributions at higher orders of $\alpha_s$ where one gluon splits into a $b\bar b$ pair, giving $bg\to H^\pm W^\mp b$. These contributions should in principle be matched to the $b\bar b\to H^\pm W^\mp$ process, which would yield an effective QCD correction factor slightly less than one~\cite{BarrientosBendezu:1998gd}. For simplicity we ignore this factor throughout this work.
If the charged Higgs boson is light enough ($\MHp<m_t$), there is an additional contribution to $H^\pm W^\mp$ production through top quark decays $t\to b H^+$. When $\hat s$ is close to the $t{\bar t}$ threshold $\hat s\sim 4m_t^2$, on-shell top quarks can therefore give an additional production channel $gg\to t{\bar t}\to H^\pm W^\mp b{\bar b} $, which is enhanced by top resonances. However, in this work we study the case when the $H^+$ boson is heavy enough to be above the threshold for production in top decays, and this extra production channel is not relevant.

The next-to-leading order (NLO) corrections to the $H^\pm W^\mp$ cross section in the MSSM are known \cite{Yang:2000yt,Hollik:2001hy,Zhao:2005mu,Gao:2007wz,Rauch:2008fy,Dao:2010nu}, but for the present study the leading order contributions suffice since we are mainly interested in comparing the NMSSM to the MSSM.
We will, however, account for the most important higher order
contributions from running quark masses, loop corrections to Higgs masses and mixings, and including (mass-dependent) widths of Higgs bosons appearing in $s$-channel
propagators. The treatment of these effects is described in further detail below.

Consider first the $b\bar{b}\to \HW$ contribution, which to the
leading order is given by the tree-level diagrams in
Fig.~\ref{fig:bbHW-diag}. At the parton-level, this contribution
typically dominates over the gluonic one considered below.
The corresponding parton-level cross section has the same form as in the
MSSM case \cite{BarrientosBendezu:1998gd},
\begin{align}\label{bbCS}
\frac{\mathrm{d}\hat\sigma}{\mathrm{d}\hat t}(b\bar{b}\to H^+W^-)=&\frac{\GF^2}{24\pi \hat
s}\Bigg\{\frac{m_b^2}{2}\lambda(\hat s,m_W^2,\MHpsq)\Big(|{\cal
S}_b(\hat s)|^2+|{\cal P}_b(\hat
s)|^2\Big)\\
&+\frac{m_b^2\tan\beta}{\hat t-m_t^2}\Big(m_W^2\MHpsq-\hat s p_{\perp}^2 -
\hat t^2\Big)\mathrm{Re}\Bigl[{\cal S}_b(\hat s)-{\cal P}_b(\hat s)\Bigr] \nn \\
&+\frac{1}{(\hat t-m_t^2)^2}\Bigl[m_t^4\cot^2\beta\Big(2m_W^2+p_{\perp}^2\Big)+m_b^2\tan^2\beta\Big(2m_W^2
p_{\perp}^2+\hat t^2\Big)\Bigr]\Bigg\}\,, \nn
\end{align}
where $\GF$ is the Fermi constant, $\shat$, $\that$, $\uhat$ are the usual Mandelstam variables, $p_{\perp}$ is the transverse momentum of the $W$ boson in the
$b\bar{b}$ c.m.~frame, and $\lambda(x,y,z)=x^2+y^2+z^2-2(xy+yz+zx)$ is the K\"all\'en
function. The first line in Eq.~(\ref{bbCS})
represents the $s$-channel resonance contribution, the last line
corresponds to the non-resonant top quark exchange in the $t$-channel, and the second line contains the interference term. 

The functions ${\cal S}_q$ and ${\cal P}_q$ contain the propagators
and relative couplings for the neutral Higgs bosons to quark flavor $q$. In
Eq.~(\ref{bbCS}) only the $b$ quark contribution is needed, but for
the $gg$ contribution discussed below we need also the corresponding expressions for the top quark. These functions are defined as
\begin{align}
{\cal S}_t(\hat s) &= \f{1}{\sin\beta} \sum_{i=1,2,3}\f{S_{i2}\,(S_{i2}\cos\beta-S_{i1}\sin\beta)}{\hat s - m_{H_i}^2 + \mathrm{i} m_{H_i} \Gamma_{H_i}}\,, \nn \\
{\cal S}_b(\hat s) &= \f{1}{\cos\beta} \sum_{i=1,2,3}\f{S_{i1}\,(S_{i2}\cos\beta-S_{i1}\sin\beta)}{\hat s - m_{H_i}^2 + \mathrm{i} m_{H_i} \Gamma_{H_i}}\,, \nn  \\
{\cal P}_t(\hat s) &= \f{1}{\sin\beta} \sum_{i=1,2}\f{P_{i2}\,(P_{i2}\cos\beta-P_{i1}\sin\beta)}{\hat s - m_{A_i}^2 + \mathrm{i} m_{A_i} \Gamma_{A_i}}\,, \nn \\
 {\cal P}_b(\hat s) &= -\f{1}{\cos\beta} \sum_{i=1,2}\f{P_{i1}\,(P_{i2}\cos\beta-P_{i1}\sin\beta)}{\hat s - m_{A_i}^2 + \mathrm{i} m_{A_i} \Gamma_{A_i}}\,, \label{NW-facts}
\end{align}
where $\Gamma_{H_i}$ and $\Gamma_{A_i}$ are the total (mass-dependent) decay widths
of the $H_i$ and $A_i$ bosons, respectively.
We have obtained the expressions given in Eq.~(\ref{NW-facts}) by modifying
the ${\cal S},\,{\cal P}$ functions given in \cite{BarrientosBendezu:1998gd} with the appropriate Yukawa
couplings for the NMSSM case.
We neglect the Yukawa couplings of the first- and second-generation quarks, as their
contributions to the amplitude are negligibly small. If the masses of two (or more) neutral Higgs bosons with the same CP properties become degenerate, then the approximation used in Eq.~(\ref{NW-facts}) breaks down,
and one has to take into account Higgs mixing effects (see e.g.\ Refs.~\cite{Pilaftsis:1997dr,Ellis:2004fs}). For the NMSSM scenarios we shall consider below, the masses will however always be such that Eq.~(\ref{NW-facts}) remain valid.

Let us now turn to the $gg$ contribution. In analogy to the MSSM case
\cite{BarrientosBendezu:1998gd,BarrientosBendezu:1999vd,BarrientosBendezu:2000tu,Brein:2000cv}, the resonant amplitude of the $gg\to H_{1,2,3},\,A_{1,2}\to \hp W^\mp$ subprocess from quark loops is given by the sum of all triangle diagrams of the type shown as the first diagram (upper left) in Fig.~\ref{fig:ggHW-diag}. This contribution can be written as
\begin{align}
 V^{\Delta}_{\lambda_W} =& \f{\sqrt{2}}{\pi} \alpha_s(\muR) \GF m_W \epsilon_{\gamma}^*(p_W)(q_1+q_2)^\gamma \nn \\
&\times \epsilon_\mu^c(q_1) \epsilon_\nu^c(q_2) \Bigg[ \Big( q_2^\mu
q_1^\nu-\f{\hat s}{2}g^{\mu\nu}\Big) \Sigma(\hat s) + \si
\epsilon^{\mu\nu\rho\sigma} q_{1\rho} q_{2\sigma} \Pi(\hat s) \Bigg]
, \label{Vhard}
\end{align}
where $\alpha_s(\muR)$ is the strong coupling evaluated at the
renormalization scale $\muR$, $\epsilon_{\gamma}^*$ is the polarization
vector of the $W$ boson with momentum $p_W$ and helicity
$\lambda_W$, and $\epsilon_{\mu,\nu}^c$ are the polarization vectors
of the gluons with momenta $q_{1,2}$. These are summed over the
color index $c$. The functions $\Sigma$ and $\Pi$ come from the loop
integration and correspond to neutral CP-even $H_{1,2,3}$ exchanges
($\Sigma$) and neutral CP-odd $A_{1,2}$ exchanges ($\Pi$) in the
$s$-channel. They are given by
\begin{align}
 \Sigma(\hat s) &= \sum_{q} {\cal S}_{q}(\hat s) S\left(\f{\hat s+\si\epsilon}{4m_{q}^2}\right), \\
 \Pi(\hat s) &= \sum_{q} {\cal P}_{q}(\hat s) P\left(\f{\hat s+\si\epsilon}{4m_{q}^2}\right) ,
\end{align}
where the sums run over all quark flavors $q$ in the triangle loops, and the functions
\begin{align}
 S(r) &= \f{1}{r}\left[ 1-\left(1-\f{1}{r}\right) \text{arcsinh}^2\sqrt{-r} \right], \\
 P(r) &= -\f{1}{r} \text{arcsinh}^2\sqrt{-r},
\end{align}
must be continued analytically for three regions in $r$, such that
for $r\le 0, 0<r\le 1$, or $r>1$, $\text{arcsinh}\sqrt{-r}$ must be represented by
$\text{arcsinh}\sqrt{-r}, -\si \arcsin\sqrt{r}$, or
$\text{arccosh}\sqrt{r} -\si\pi/2$, respectively. The contribution to the
parton-level cross section is then given by
\begin{equation}
\frac{\mathrm{d}\hat\sigma^{\Delta}}{\mathrm{d}\that}(gg\to H^+W^-)=\frac{\alphas^2(\muR)\GF^2}{2048\pi^3}\lambda(\hat
s,m_W^2,\MHpsq)\Big(|\Sigma(\hat s)|^2+|\Pi(\hat s)|^2\Big)\, .
\label{eq:ggHW}
\end{equation}
Due to
Bose symmetry, the $gg\to H^{\pm}W^{\mp}$ cross section is symmetric
with respect to $\hat t\leftrightarrow \hat u$ interchange.
Additionally, since we only consider the CP-invariant case,
the cross sections for the $gg\to H^+W^-$ and $gg\to H^-W^+$ channels coincide.

In our numerical calculations we take all possible quark and
squark loop contributions into account from both triangle and box diagrams. For simplicity, we show here only the formulas for quark triangles, and do not list the complicated expressions for either the boxes or the diagrams with squark loops, schematically shown in Fig.~\ref{fig:ggHW-diag}. The full result also includes interference between these different contributions.  We have checked our numerical results in the MSSM limit (which will be described below) against previous results from the literature \cite{BarrientosBendezu:1998gd,Dao:2010nu}.

Formally, the leading order contributions (as given by Eqs.~(\ref{NW-facts})--(\ref{eq:ggHW}) above) contain tree-level masses and couplings. As advocated previously (see \cite{Dao:2010nu} and references therein), higher order QCD and electroweak corrections can significantly affect MSSM observables. In particular the bottom Yukawa coupling is subject to large quantum corrections in the MSSM---as well as in the
NMSSM---and these need to be taken into account properly. 
For this purpose, we follow the general recipe given in \cite{Dao:2010nu}. To take into account the large (SM) QCD corrections to the leading-order result, we use the QCD
running $b$ quark mass $m_b=m_b^{\DRbar}(\muR)$. At
two-loop order it is given by \cite{Avdeev:1997sz}
\begin{eqnarray}
m_b^{\DRbar}(\mu_R)=m_b^{\MSbar}(\mu_R)
\Big[1-\frac{\alphas}{3\pi}-\frac{\alphas^2}{144\pi^2}(73-3n)\Big],
\end{eqnarray}
where $n$ is the number of active quark flavors and
$m_b^{\MSbar}(\mu_R)$ is the standard $\MSbar$ running
mass (we use $m_b^{\MSbar}(m_b)=4.2\GeV$ as input). Then, including the $\tan\beta$-enhanced supersymmetric QCD (SQCD) and electroweak (SEW) corrections \cite{Hall:1993gn} by a straightforward generalization of the MSSM results \cite{Carena:1998gk}, we obtain the following effective
bottom-Higgs couplings:
\begin{align}
\lambda^{\mathrm{eff}}_{b{\bar b}H_i}&=-\si\frac{
m_b^{\DRbar}}{\sqrt{2}\, v \cos\beta}\frac{S_{i1}}{1+\Delta_b}\left(1+\Delta_b \frac{S_{i2}}{S_{i1}\tan\beta
}\right),\label{deltab}\quad
i=1,2,3 \\
\lambda^{\mathrm{eff}}_{b{\bar
b}A_k}&=\frac{ m_b^{\DRbar}}{\sqrt{2}\, v \cos\beta}\frac{P_{k1}}{{1+\Delta_b}}
\left(1+\Delta_b \frac{P_{i2}}{P_{i1}\tan\beta
}\right),\quad
k=1,2
\end{align}
In a similar manner we also include the relevant corrections to the $H^+tb$ vertex \cite{Carena:1999py}. These so-called $\Delta_b$ corrections consist of two dominating parts, $\Delta_b=\Delta_b^{\mathrm{SQCD}}+\Delta_b^{\mathrm{SEW}}$, absorbing the leading SQCD and SEW corrections. In our case, the latter is dominated by the Higgsino-stop contribution $\Delta_b^{\mathrm{SEW}}\simeq \Delta_b^{\tilde{H}\tilde{t}}$. In complete analogy to the MSSM case we therefore 
have $\Delta_b \simeq \Delta_b^{\mathrm{SQCD}}+
\Delta_b^{\tilde{H}\tilde{t}}$, where
\begin{eqnarray*}
&&\Delta_b^{\mathrm{SQCD}}=\frac{2\alphas(Q)}{3\pi} m_{\tilde{g}}\,\mueff
\tan\beta\,
I(m_{\tilde{b}_1}^2,m_{\tilde{b}_2}^2,m_{\tilde{g}}^2),\quad
Q=(m_{\tilde{b}_1}+m_{\tilde{b}_2}+m_{\tilde{g}})/3,\\
&&\Delta_b^{\tilde{H}\tilde{t}}=\frac{m_t^2}{16\pi^2v^2
\sin^2\beta}
A_t\,\mueff\tan\beta\,I(m_{\tilde{t}_1}^2,m_{\tilde{t}_2}^2,|\mueff|^2),
\end{eqnarray*}
and finally
\begin{eqnarray*}
I(a,b,c)=-\frac{1}{(a-b)(b-c)(c-a)}\left(ab\ln\frac{a}{b}+bc\ln\frac{b}{c}+ca\ln\frac{c}{a}\right).
\end{eqnarray*}
Here $m_{\tilde{g}}$ denotes the gluino mass, and $m_{\tilde{b}_i},\,m_{\tilde{t}_i}$ ($i=1,2$) the sbottom and stop masses. From these expressions it is clear that the $\Delta_b$ corrections could become large for either large values of $\mueff$ and/or large $\tan\beta$.

We shall refer below to the calculation at leading order, including the improvements discussed here (higher order corrections to Higgs masses and mixing, the Higgs widths in the $s$-channel Higgs propagators, the running $m_b$, and the SUSY corrections to the bottom Yukawa couplings), as
the {\it improved Born approximation}.

To calculate the cross sections and perform the numerical computations, we have modified and extended the
MSSM model file \cite{Hahn:2001rv} of \FA{} \cite{Feynarts}
to contain the relevant NMSSM couplings and the necessary steps to use the improved Born approximation
as discussed above. The parton-level amplitudes have
been computed with \FC~\cite{FormCalc} and integrated
numerically. For the evaluation of the scalar master tree- and
four-point integrals in the gluon contribution we have used the \LT{} library \cite{FormCalc}. The Higgs mass spectra, mixing, couplings and decay widths
have been calculated using \NMT{}~\cite{Ellwanger:2004xm,Ellwanger:2005dv}.

\section{Results}
\subsection{NMSSM parameter dependence and benchmark scenarios}
As a first step, we investigate the impact on the parton-level cross sections $\sigbbhw$ and
$\siggghw$ of varying the NMSSM
parameters. This information will be useful for defining the
benchmark scenarios we are going to study in more detail below. We
start from a generic NMSSM scenario with parameter values chosen as
follows (this is what we will below refer to as Scenario A):
\begin{equation}
\begin{aligned}
\lambda&=0.25, \quad \kappa=0.25,\quad A_\lambda=-235\GeV,\\
 A_\kappa&=-150\GeV,\quad \tan\beta=10,
\end{aligned}
\label{eq:params}
\end{equation}
and perform variations around these values. For each parameter point
the partonic cross sections are evaluated as a function of
$\sqrt{\hat{s}}$, using the improved Born approximation as described
in Section~\ref{sect:theory}.

\begin{figure}
\includegraphics[width=0.48\columnwidth]{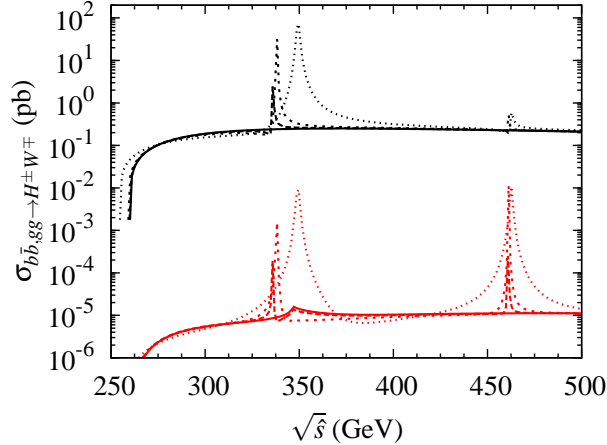}
\caption{Parton-level cross sections for $H^\pm W^\mp$ production by
$b\bar{b}$ (black) and $gg$ (red) initial states for
$\lambda$=$\kappa=0.25$ (dotted), $0.1$ (short-dashed), $0.05$
(long-dashed), and in the MSSM limit $\lambda=\kappa=10^{-10}$
(solid). The remaining NMSSM parameters are fixed according to the
description given in the text.} \label{fig:mssmlim}
\end{figure}
To be able to study genuine NMSSM effects on the $H^\pm W^\mp$
process, we first want to compare our results to those obtained in
the MSSM limit. We therefore look at the behavior of the partonic
cross sections during the gradual transition from the NMSSM point
defined by the parameter set given by Eq.~\eqref{eq:params} to the
corresponding MSSM limit.\footnote{We remind the reader that the
MSSM limit is defined by taking $\lambda,\,\kappa\to 0$, while
keeping the ratio $\kappa/\lambda$ and all other parameters fixed to
their respective values.} The results are shown in \reffig{mssmlim}. A striking difference between the NMSSM
and the MSSM is the presence of resonant enhancement of the partonic
cross sections in the NMSSM. These resonances can be attributed to the heavy
neutral Higgs poles $\hat{s}=m^2_{H_3}$ and $\hat{s}=m^2_{A_2}$,
which for the default parameters have masses $m_{H_3}=462.6\GeV$ and
$m_{A_2}=349.3\GeV$ (see Table \ref{tab:benchmarks} below). Since in
the MSSM it is generally true that $m_{H,A}<\MHp$, these
resonant contributions vanish in the MSSM limit. They are therefore
an inherent feature of the NMSSM. As we will show below, the
resonant contributions are sensitive to the NMSSM parameters, and
can give important contributions to the rate for $pp\to \HW$
production. They are therefore interesting to study as a means of
discriminating between the two models.

\begin{figure}
\includegraphics[width=0.45\columnwidth]{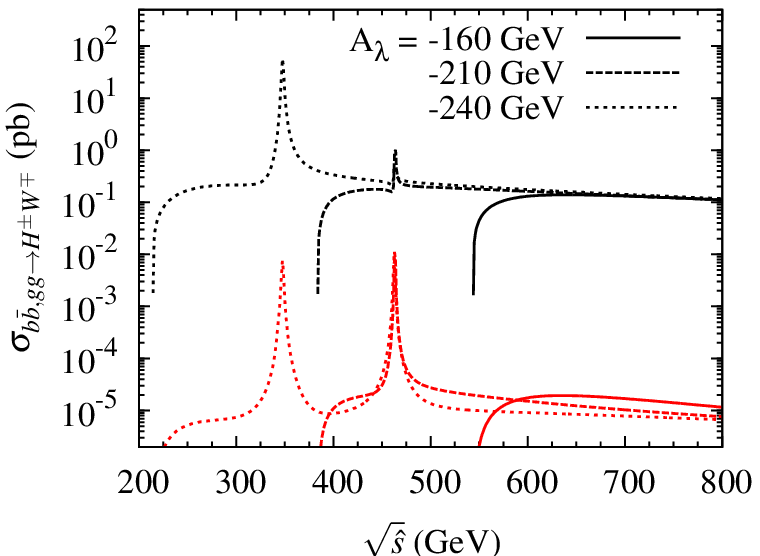}\;
\includegraphics[width=0.45\columnwidth]{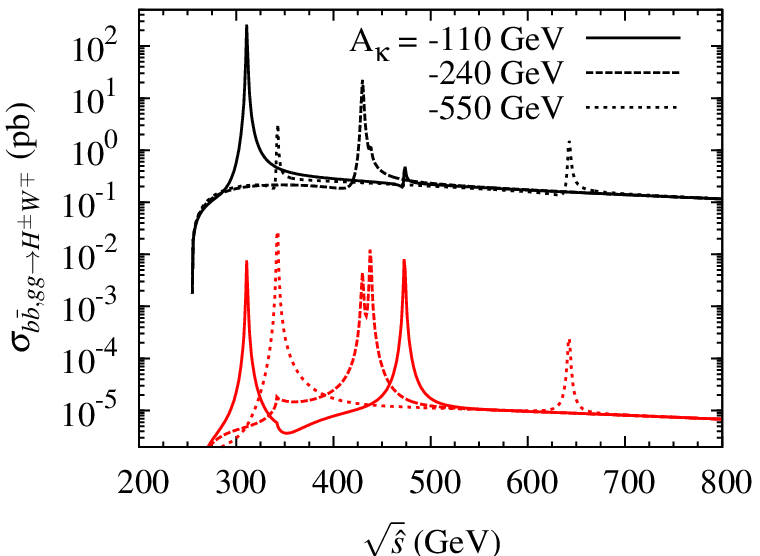}\\\vspace{0.25cm}
\includegraphics[width=0.45\columnwidth]{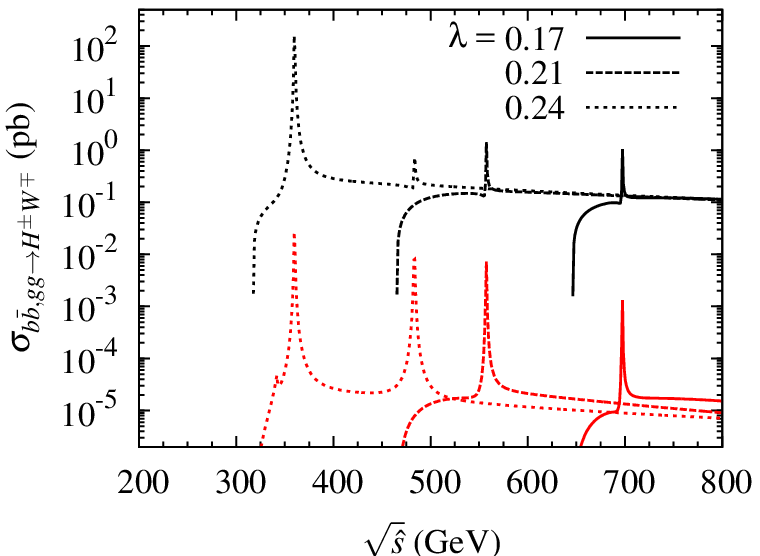}\;
\includegraphics[width=0.45\columnwidth]{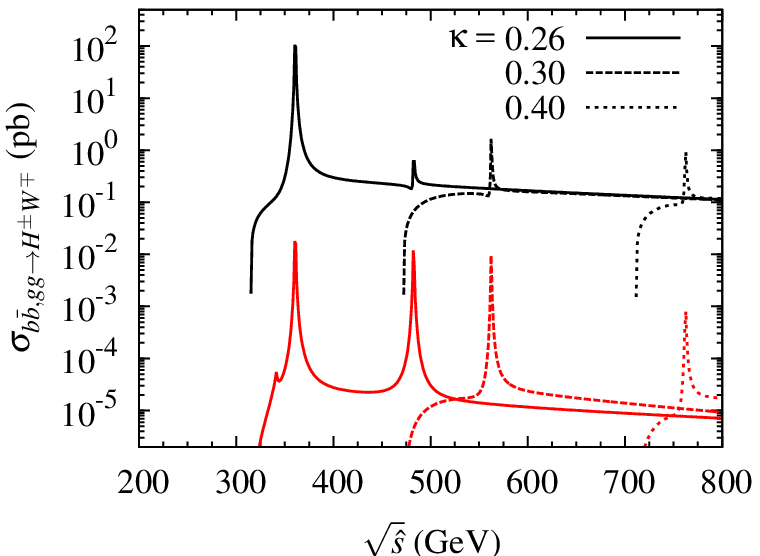}\\\vspace{0.25cm}
\includegraphics[width=0.45\columnwidth]{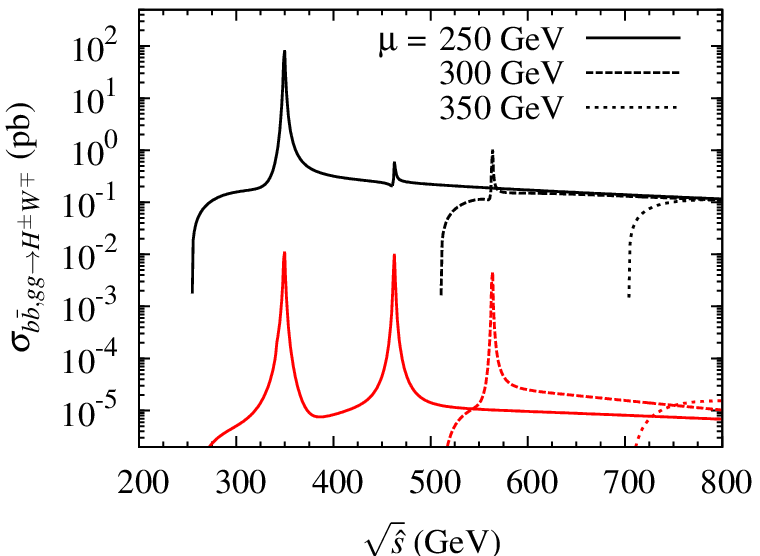}\;
\includegraphics[width=0.45\columnwidth]{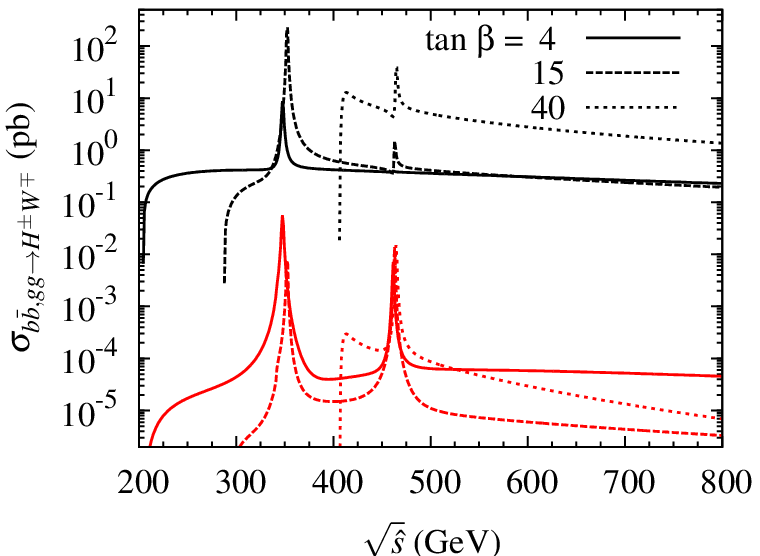}
\caption{Parton-level cross sections for $H^\pm W^\mp$ production by $b\bar{b}$ (black) and $gg$ (red) initial states versus $\sqrt{\hat{s}}$ for different values of the NMSSM parameters, as indicated in the plots. The remaining NMSSM parameters are fixed according to the description in the text.}
\label{fig:varparam}
\end{figure}
Having established the presence of resonances as a potentially
important difference for $H^\pm W^\mp$ production between the MSSM
and NMSSM, we now proceed to study how the characteristics of these
resonances are affected by variation of the NMSSM parameters. The
results are shown in \reffig{varparam}. Starting from the
upper left plot, we first consider different values of $A_\lambda$.
This has an immediate effect on the threshold through the linear
dependence of $\MHpsq$ on $A_\lambda$ (see
Eq.~(\ref{eq:mhp2})). It can also be seen that a change in
$A_\lambda$ affects the presence of a resonance peak in the cross
section. Qualitatively this can be understood from the fact that a
charged Higgs boson with lower $\MHp$ can be produced from a
(nearly) singlet scalar (pseudoscalar) of fixed mass, while this
possibility disappears for a particular mass as $\MHp$ is
increased. Further details on this point will be given below. Moving
to the next plot, we see that a variation of $A_\kappa$ affects the
cross sections in quite a different manner; it affects the relative
positions and sizes of the resonance peaks. Effectively, when going
from low $|A_\kappa|$ to higher values the two peaks, which at first
are well separated, first meet and then get separated again, having
changed positions. Since it also affects the relative size of the
peaks, this signals that the mixture of the neutral Higgs resonances
changes with modified $A_\kappa$. The two plots in the second row
show the variation with $\lambda$ and $\kappa$, respectively. In
addition to the similar shift in threshold as demonstrated for the
$A_\lambda$ case, we also see that these cross sections appear to
depend on these parameters in the same combination as appears in
$\MHp$ for fixed $\mueff$, that is through the ratio
$\kappa/\lambda$. On the last row of \reffig{varparam} we
show the variation of the cross sections with $\mueff$. This gives a
very similar effect as choosing different values for $A_\lambda$,
which motivates us to consider a fixed value $\mueff=250\GeV$ in the
following and instead use variation of $A_\lambda$ to control the
value of $\MHp$. The final plot shows the effect of a
$\tan\beta$ variation. Also this variable enters the determination
of $\MHp$ (and thereby shifts the threshold), although in a
more indirect way than $\mueff$ or $A_\lambda$. As can be seen from the
figure, changing $\tan\beta$ also has a drastic effect on the
absolute normalization of the cross sections and the width of the
resonances. This comes mainly from the coupling of the charged Higgs
boson to fermions of the third generation. For equivalent kinematic
configurations, we can therefore expect enhancements of the cross
sections either for small or large values of $\tan\beta$.

\begin{table}[t!]
\begin{center}
\begin{tabular}{|c||ccccc|} \hline
                    &  \multicolumn{5}{c|}{Scenario}\\
 Parameter              &     A        &      B        &    C      &    D       &     E         \\
  \hline
 $A_{\lambda}$ (GeV)    &  $-235$      &   $-235$      &  $-235$   &  $-185$    &  $-243$            \\
 $A_{\kappa}$ (GeV)     &  $-150$      &   $-250$      &  $-400$   &  $-150$    &  $-150$            \\
 $\lambda$              &  $0.25$      &   $0.25$      &  $0.25$   &  $0.5$     &  $0.25$            \\
 $\kappa$               &  $0.25$      &   $0.25$      &  $0.25$   &  $0.5$     &  $0.25$            \\
 $\tan \beta$           &    $10$      &   $10$        &   $10$    &  $2.2$         &  $40$              \\
 \hline
 \multicolumn{6}{|c|}{Higgs mass spectrum (GeV)} \\
 \hline
   $\MHp$            & $174.3$      &   $174.3$       &   $174.3$  &  $195.3$  &  $171.7$            \\ \hline
   $m_{H_1}$            & $118.4$      &   $117.4$       &   $115.0$  &  $114.6$  &  $120.3$            \\
   $m_{H_2}$            & $173.5$      &   $174.1$       &   $174.3$  &  $203.6$  &  $246.0$            \\
   $m_{H_3}$            & $462.6$       &   $435.3$      &   $391.1$ &   $459.6$  &  $463.3$            \\  \hline
   $m_{A_1}$         & $139.0 $       &    $156.4$       &   $165.4$      &  $92.0$  &  $213.2$           \\
   $m_{A_2}$         & $349.3$ &     $438.2$  &   $549.2$ &  $383.4$  &  $355.7$            \\  \hline
 \multicolumn{6}{|c|}{Singlet elements of $H_3,\,A_2$}  \\ \hline
   $S_{3,3}$  & 0.993 &  0.991 &  0.986   &  $0.988$  &  $0.992$            \\
   $P_{2,3}$  & 0.945 &  0.981 &  0.993   &  $0.875$ &   $0.897$            \\  \hline
\end{tabular} \end{center}
\caption{Selected NMSSM benchmark scenarios, the corresponding Higgs
mass spectrum, and singlet elements $S_{3,3}$, $P_{2,3}$ of the
Higgs mixing matrices for the neutral heavy Higgs bosons.}
\label{tab:benchmarks}
\end{table}
Based on these parameter variations we define five benchmark
scenarios for further study (called scenarios A--E). The scenarios
are selected to capture the different features of the partonic cross
sections discussed above, and the values for the input parameters in
the five benchmarks are shown in Table~\ref{tab:benchmarks}. This
table also gives the resulting Higgs mass spectrum, and the
$S_{3,3}$, $P_{2,3}$ elements of the Higgs mixing matrices which
give the singlet fractions of the heavy Higgs bosons $H_3$ and
$A_2$. They will be relevant for the discussion of the resonant
contributions below.

Table~\ref{tab:benchmarks} shows some features which are common to
all the scenarios. As a general strategy we choose the parameters to
obtain a rather low $\MHp$ in all scenarios (but still keeping
$\MHp>m_t$). For compatibility with LEP constraints
\cite{Barate:2003sz},  we make sure that the lightest CP-even Higgs
mass $m_{H_1}>114.4\GeV$.\footnote{Applying this limit to the NMSSM
in general is a very conservative approach, since it strictly
speaking only applies to a Higgs with SM-like couplings. Specific
NMSSM scenarios admit $m_{H_1}\ll 114.4\GeV$ without being in
conflict with experimental data. However, in this work the selected
scenarios correspond to the case where $H_1$ is SM-like, and
therefore the SM limit applies.} The scenarios are also compatible
with the limits from direct Higgs searches at the Tevatron and the
LHC implemented in \HB~v.2.1.0~\cite{Bechtle:2011sb}.
Going into the specifics of the individual scenarios, the only
difference between the definitions of scenarios A, B, and C is the
value for $A_\kappa$. This variation leads to three different
hierarchies for $m_{H_3}$ and $m_{A_2}$, as can be read off from the
table. Scenarios D and E are both mostly similar to Scenario A in
terms of the heavy Higgs mass structure. Here we instead consider two
``extreme'' cases of low (D) and high (E) values for $\tan\beta$.

\subsection{Parton-level cross sections}
\label{sect:parton}
\begin{figure}
\includegraphics[width=0.45\columnwidth]{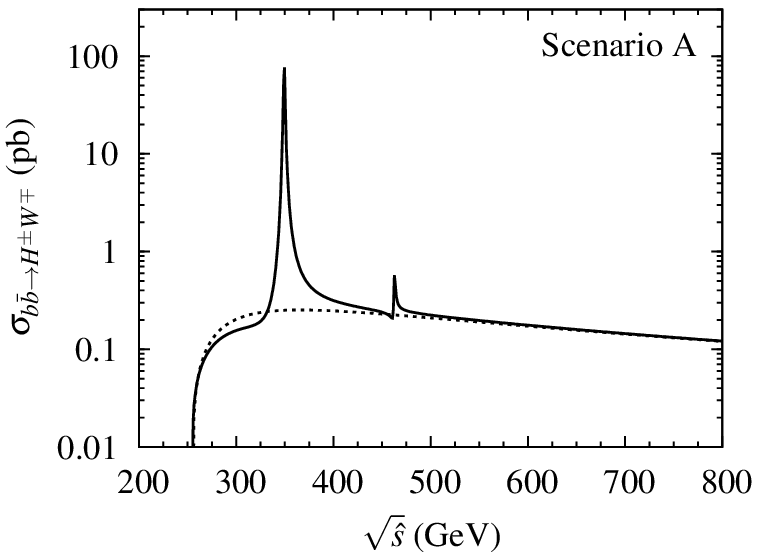}
\includegraphics[width=0.45\columnwidth]{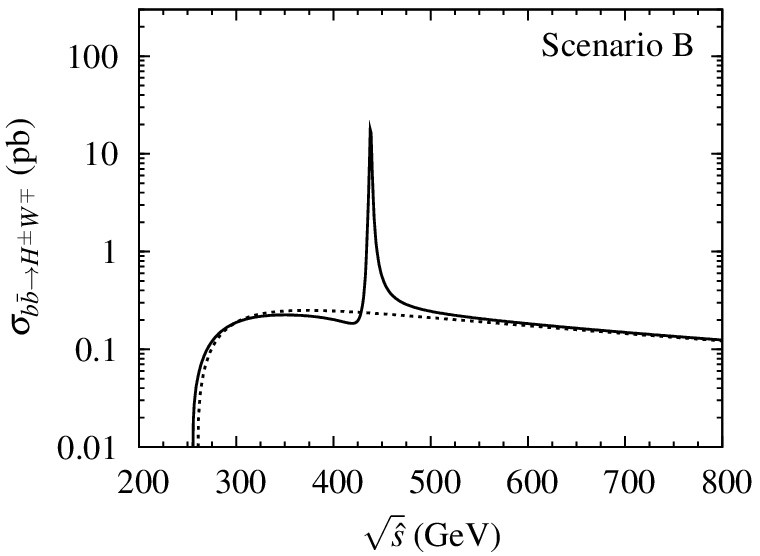}\\\vspace{0.25cm}
\includegraphics[width=0.45\columnwidth]{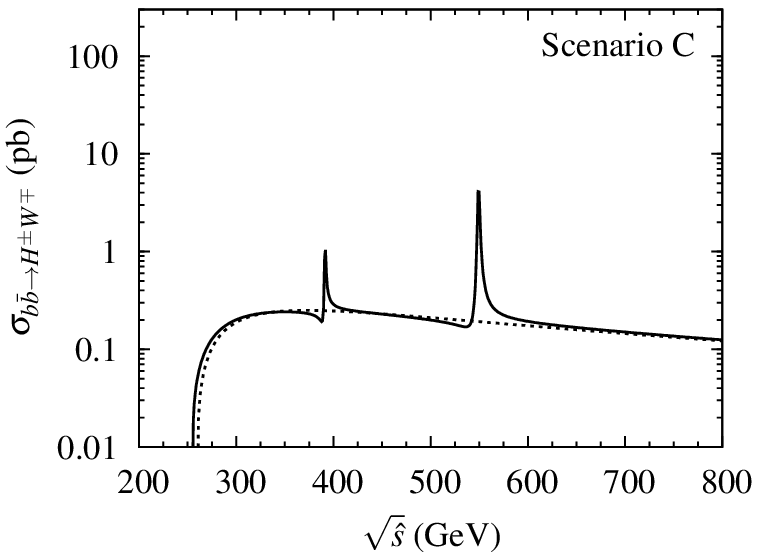}
\includegraphics[width=0.45\columnwidth]{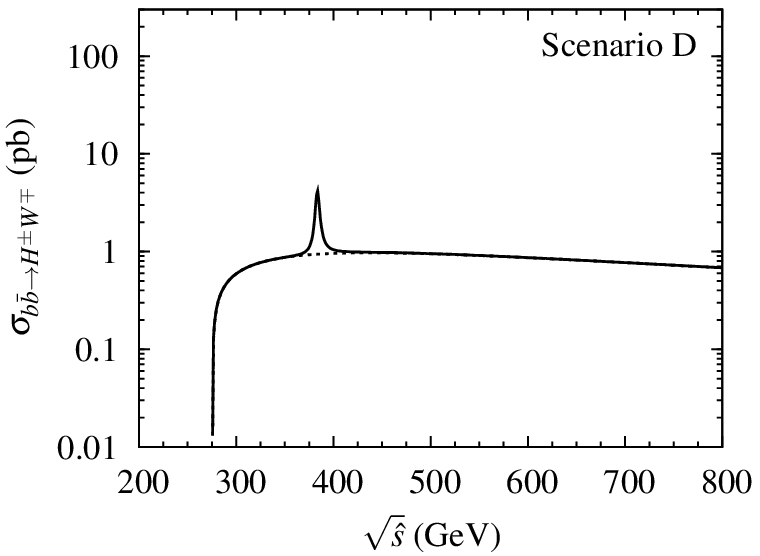}\\\vspace{0.25cm}
\includegraphics[width=0.45\columnwidth]{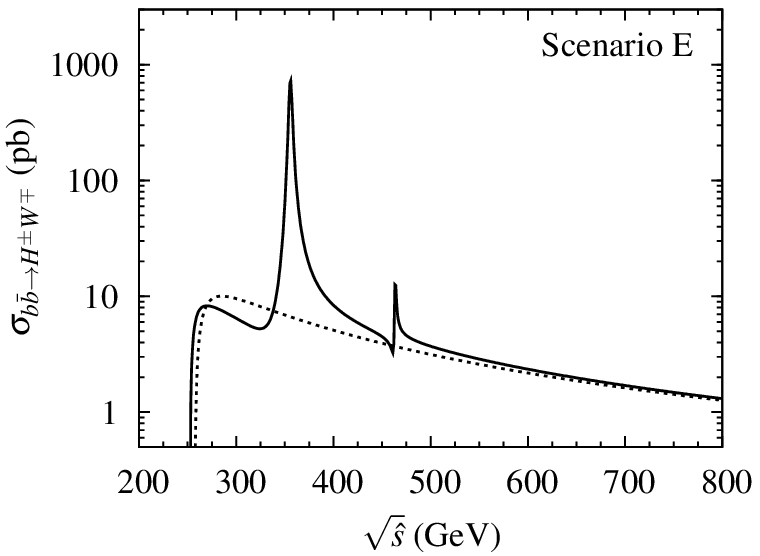}
   \caption{Partonic cross sections $\hat{\sigma}_{b\bar{b}\to H^\pm W^\mp}$ for NMSSM benchmark scenarios A--E (solid lines), and the corresponding results in the MSSM limit (dashed).}
 \label{fig:bbHW}
\end{figure}

We now proceed to study associated $H^\pm W^\mp$ production in the NMSSM, making use of the benchmark scenarios defined in the previous section. The parton-level cross sections are evaluated as described in Section~\ref{sect:theory}, and the results for $\sigbbhw$ are shown in \reffig{bbHW}.

\begin{figure}[t!]
\includegraphics[width=0.45\columnwidth]{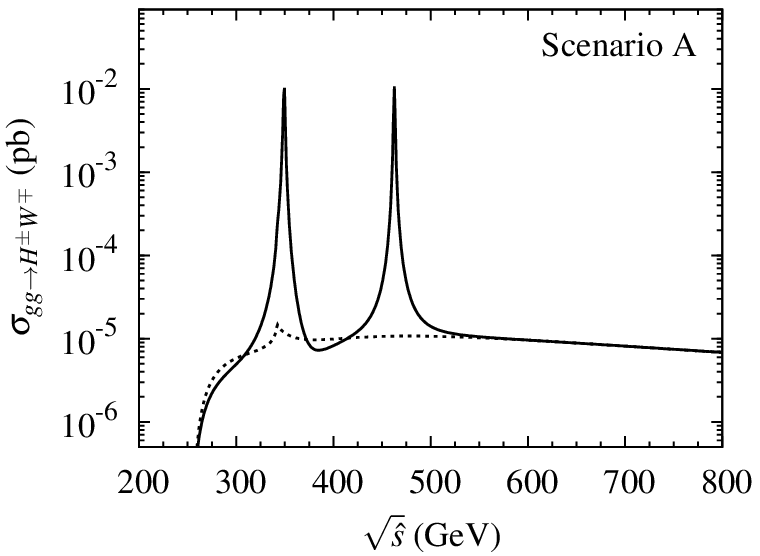}
\includegraphics[width=0.45\columnwidth]{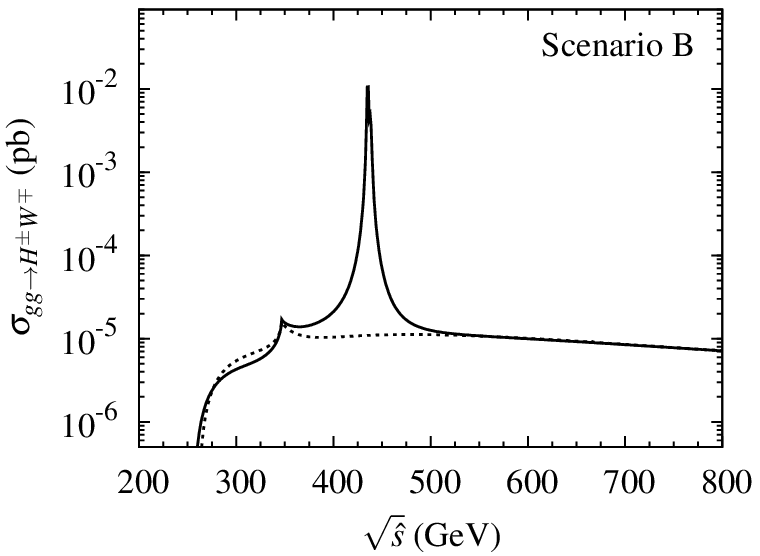}\\\vspace{0.25cm}
\includegraphics[width=0.45\columnwidth]{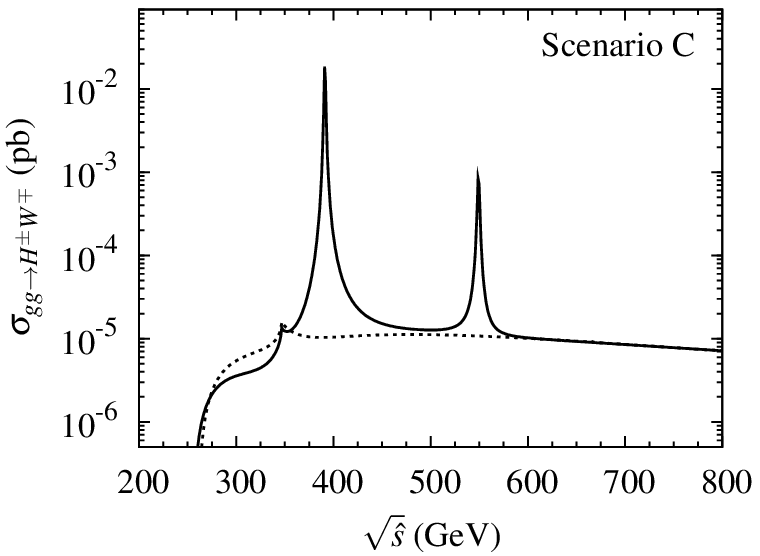}
\includegraphics[width=0.45\columnwidth]{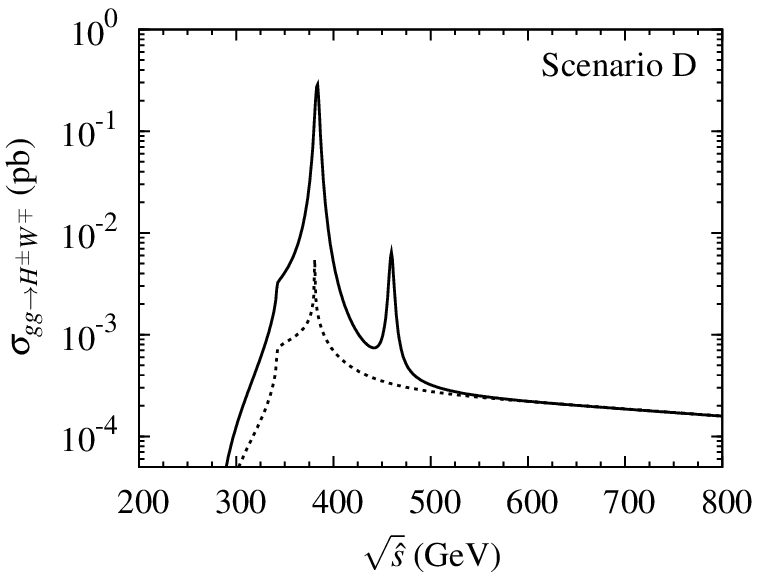}\\\vspace{0.25cm}
\includegraphics[width=0.45\columnwidth]{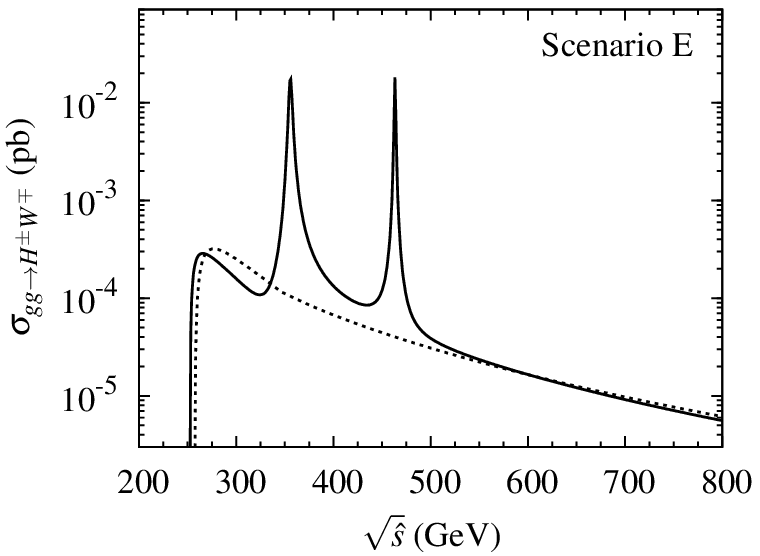}
   \caption{Partonic cross sections $\hat{\sigma}_{gg\to H^\pm W^\mp}$ for NMSSM benchmark scenarios A--E (solid lines), and the corresponding results in the MSSM limit (dashed).}
 \label{fig:ggHW}
\end{figure}

In this figure the solid lines give the NMSSM cross sections, while
the dashed lines are the corresponding cross sections in the MSSM
limit. As already discussed above, the most striking difference is
the presence of resonances in the NMSSM case. In all the five
benchmarks the resonances are manifest as either one or two peaks of
varying size. Comparing the plots for scenarios A--C, we observe a
shift of the large peak from low energies (for Scenario A) to high
energies (Scenario C). There is also a corresponding down-shift of
the smaller peak to low energies. For Scenario B the two peaks
nearly coincide. The positions of these poles are determined by
$m_{H_3}$ and $m_{A_2}$, and we note that the largest peak
corresponds to the $A_2$ resonance. This means that even if the
singlet component of $A_2$ is larger than that of $H_3$
($|P_{2,3}|>|S_{3,3}|$), the $A_2$ couples more strongly to the
$b\bar{b}$ initial state. For Scenario D we only observe one (small)
resonance peak. This is due to the low value of $\tan\beta$,
corresponding to a reduced $A_2 b\bar{b}$ coupling. On the other
hand, we also observe a larger overall cross section in the
continuum (visible also in the MSSM limit), since the non-resonant
contribution mediated by $t$-channel top exchange increases as
$\tan\beta\ll 7$. For Scenario E (which has a high value of
$\tan\beta$) something similar is observed for the continuum, but
with an even larger enhancement of the cross section at low energies
and a more rapid drop as $\sqrt{\hat{s}}\to \infty$. In this case we
also observe two pronounced resonances (as in scenarios A and C),
but the fairly large couplings in Scenario E lead to larger
differences between the NMSSM results and the MSSM limit also for
energies away from the actual poles.

Turning now to $\siggghw$, the contribution from gluon fusion is expected to be richer than that initiated by $b$-quarks, since it involves additional non-resonant box diagrams. The interference with these can strongly affect the resulting cross section (for a study of these interference effects in the MSSM, see \cite{BarrientosBendezu:1998gd}). The cross section is again evaluated as outlined in Section~\ref{sect:theory} and the results are shown in \reffig{ggHW}. One look at this figure reveals that the gluon-initiated process has a much lower cross section compared to the $b\bar{b}$ initial state, about $3$--$4$ orders of magnitude. We can however expect this difference to be (at least partly) compensated in the hadronic cross section by the larger gluon content of the proton at intermediate and small $x$ (see below).  Compared to the $b\bar{b}$ process, we do observe a general broadening of the resonances, and larger differences between the NMSSM and the MSSM limit---the latter in particular for energy ranges between near-lying resonances, where interference can lead to either an enhanced or suppressed cross section prediction in the NMSSM compared to the MSSM. Most of the $gg$ distributions show a feature at the top pair threshold $\sqrt{\hat{s}}=2m_t$ (sometimes masked by an NMSSM resonance). Since this kinematic effect is present also in the MSSM it is not interesting for the comparison of the results between the two models.

Looking specifically at the results for scenarios A--C in \reffig{ggHW}, they display the same resonance structure as the $b\bar{b}$ case. This tells us that both the $H_3$ and $A_2$ resonances play a role also here. However the peaks are more similar in size (for Scenario A), and in the case of Scenario C we see that the low energy peak is dominating the cross section. Since this corresponds to the $H_3$ contribution, we conclude that the resonant process here is instead dominated by the $H_3 t\bar{t}$ coupling which enters the top loop contribution. The importance of the coupling to the top for lower $\tan\beta$ becomes evident for Scenario D. This scenario has a greatly enhanced cross section, both compared to scenarios A--C, and with very pronounced resonance enhancements compared to the MSSM limit. Also Scenario E benefits from the same large continuum cross section as observed for the $b\bar{b}$ contribution, but with the additionally boosted resonance contribution of $H_3$ observed generally for the $gg$ process.

\subsection{Hadron-level cross sections}
\begin{figure}
\includegraphics[width=0.5\columnwidth]{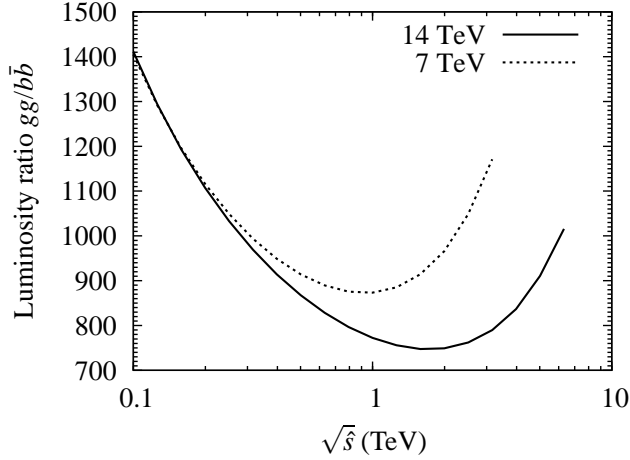}
   \caption{Ratio of parton luminosities $(\mathrm{d} L_{gg}/\mathrm{d}\hat s)/(\mathrm{d}L_{b\bar b}/\mathrm{d}\hat s)$ at 14 TeV (solid) and 7 TeV (dotted).}
 \label{fig:lumiratio}
\end{figure}

The total hadronic cross section $\sigpphw$ is obtained from the
partonic cross sections $\hat{\sigma}_{ij}$ (with $ij=b\bar{b},gg$)
by integration over the parton distribution functions (PDFs). This
can alternatively be expressed in terms of parton luminosities,
which allows studying the impact of the PDFs on the cross section.
The parton luminosities $L_{ij}$ for partons $i,j$ are defined as
\begin{equation}
 \tau \frac{\mathrm{d}L_{ij}}{\mathrm{d}\tau} = \int \mathrm{d} x_1 \, \mathrm{d}x_2 \,x_1 x_2 \delta(\tau - x_1 x_2)\Bigl\{  f_i(x_1,\muF^2)   f_j(x_2,\muF^2) +i\leftrightarrow j\Bigr\},
\label{lumi-def}
\end{equation}
where $\tau=x_1 x_2$ and $f_i(x,\muF^2)$ is the PDF for parton $i$ evaluated at the factorization
scale $\mu_F$. Note that, since we are interested in $b\bar b$ and $gg$ scattering, we
will always have $f_i(x)=f_j(x)$. The significance of $\tau$ is that the
center-of-mass energy of the partonic system is given by $\sqrt{\hat
s}=\sqrt{\tau s}$, where $\sqrt{s}=E_{\mathrm{CM}}$ is the collider energy.
Using the parton luminosities, the total cross section can be expressed as
\begin{equation}
 \sigpphw(s) = \int_0^1 \frac{\mathrm{d}\tau}{\tau} \left[ \frac{1}{s} \frac{\mathrm{d}L_{ij}}{\mathrm{d}\tau}\right] \left[\hat s \hat\sigma_{ij}(\hat s)\right].
\end{equation}
Note that the factor involving the parton luminosity has the
dimensions of a cross section; it can be used to estimate the size of
the hadronic cross section when the partonic cross section is known.
Even though the $gg$ cross sections are much
smaller than the $b\bar b$ cross sections at parton-level, at hadron-level the
much larger gluon PDFs make the $gg$ contribution competitive with
the $b\bar b$ contribution. In Fig.~\ref{fig:lumiratio} we show the
ratio of the $gg$ and $b\bar b$ luminosities at 7 TeV and at 14 TeV,
calculated with MSTW PDFs \cite{Martin:2009iq}. We see see from this figure that
the $gg/b\bar{b}$ luminosity ratio is typically a factor $1000$ at 7~TeV, and slightly smaller for 14~TeV.

\begin{figure}[b]
\includegraphics[width=0.48\columnwidth]{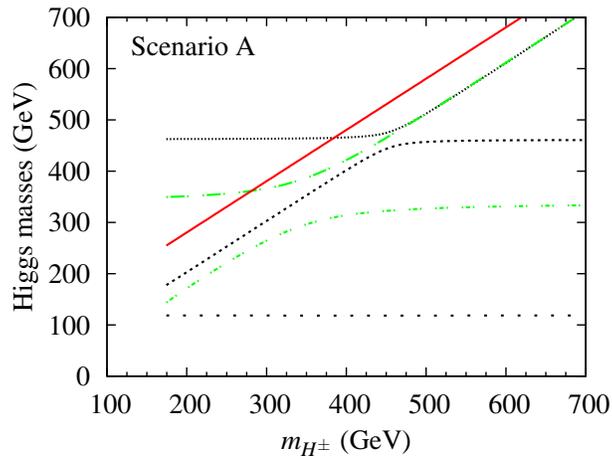}
\caption{Dependence of the neutral Higgs masses on $\MHp$ in
Scenario A with varying charged Higgs boson mass. The different
curves correspond to (in black): $m_{H_1}$ (sparse dots), $m_{H_2}$
(dotted), $m_{H_3}$ (dense dots), and in green $m_{A_1}$
(dot-dashed, short), and $m_{A_2}$ (dot-dashed, long). The solid red
line shows the threshold for associated $H^\pm W^\mp$ production.}
\label{fig:hmasses}
\end{figure}
For the calculation of the hadronic cross sections, we
modify the benchmark scenarios listed in Table~\ref{tab:benchmarks}
to allow for variation of $\MHp$. This is achieved by varying
the value of $A_\lambda$, keeping the other parameters fixed to the values given in the table. Besides changing $\MHp$, this affects
the doublet-dominated neutral Higgs bosons (which never appear resonantly), while the masses of the singlet-dominated Higgses ($H_3$ and $A_2$ for low $\MHp$) are
largely insensitive to $A_\lambda$ variations. This means in particular that
the resonance structure discussed above will remain unaltered over certain ranges for $\MHp$ (when the doublet mass is well below the singlet mass scale). This important point is
demonstrated in \reffig{hmasses}, which shows the dependence
of the five neutral Higgs masses in Scenario A on $\MHp$ (when
varying $A_\lambda$). A  qualitatively similar picture is obtained in the
other scenarios we study.

\begin{figure}[b!]
\includegraphics[width=0.45\columnwidth]{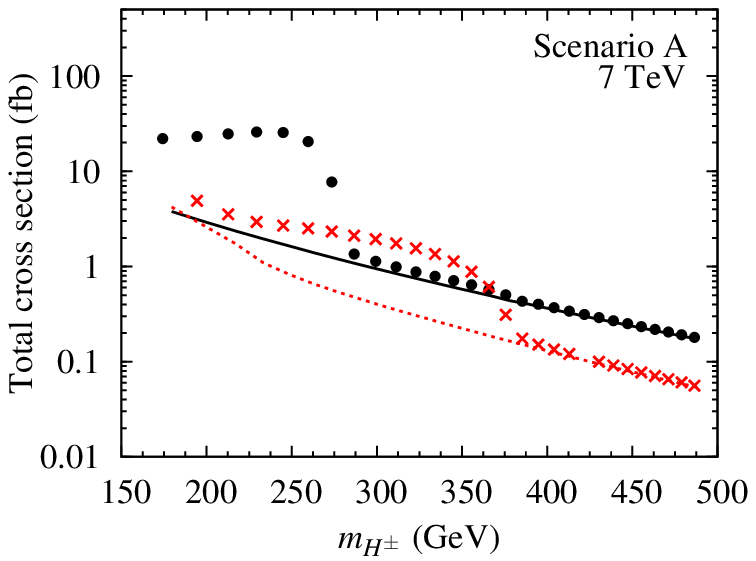}
\includegraphics[width=0.45\columnwidth]{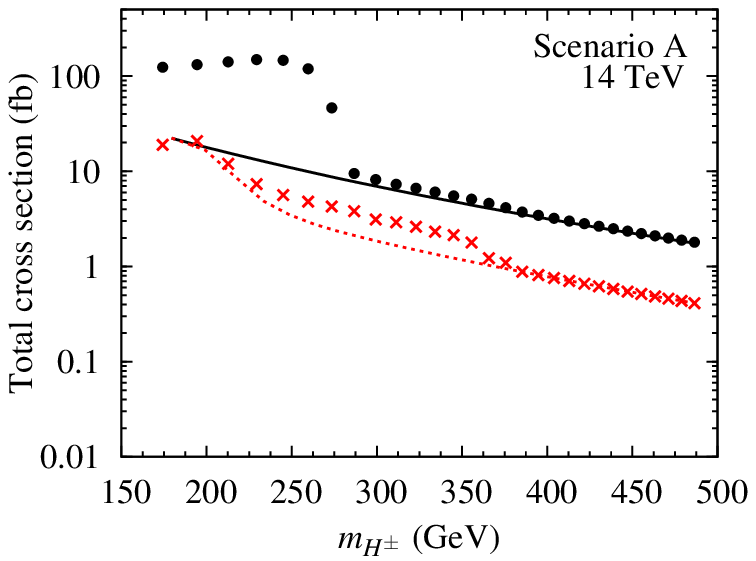}\\\vspace{0.2cm}
\includegraphics[width=0.45\columnwidth]{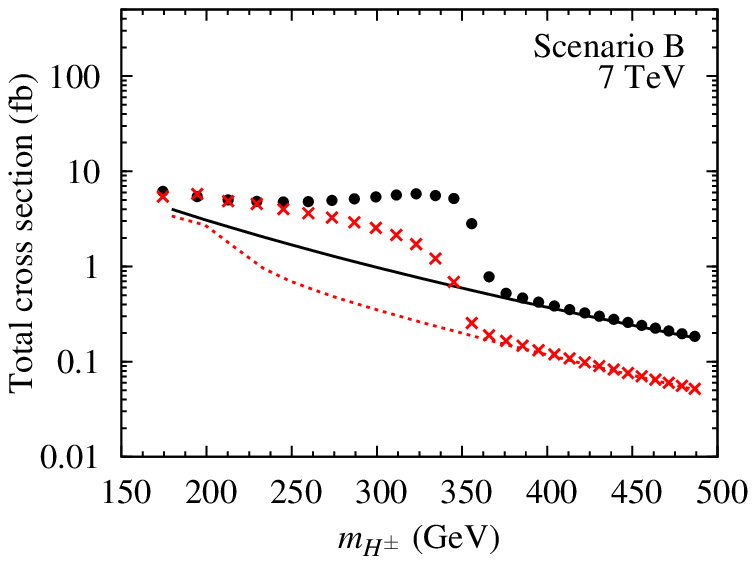}
\includegraphics[width=0.45\columnwidth]{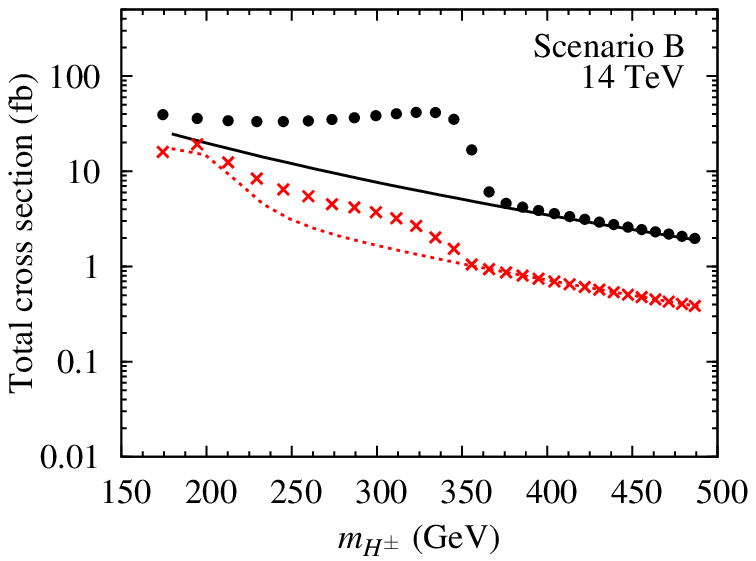}\\\vspace{0.2cm}
\includegraphics[width=0.45\columnwidth]{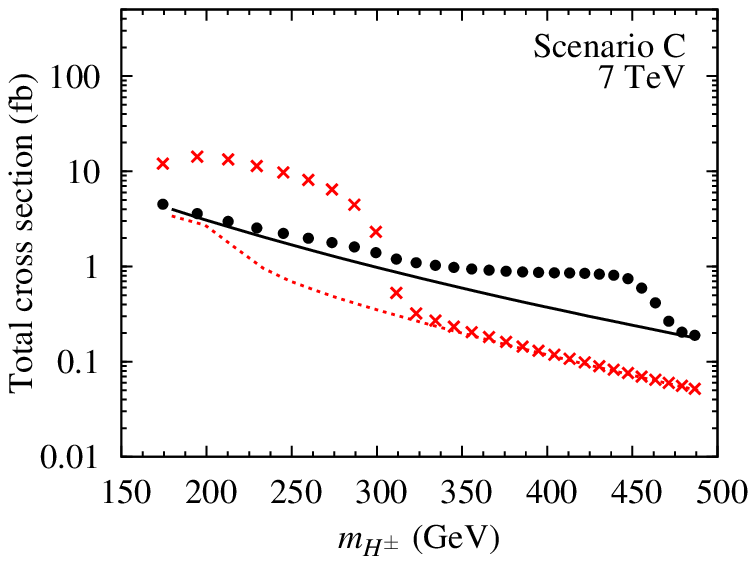}
\includegraphics[width=0.45\columnwidth]{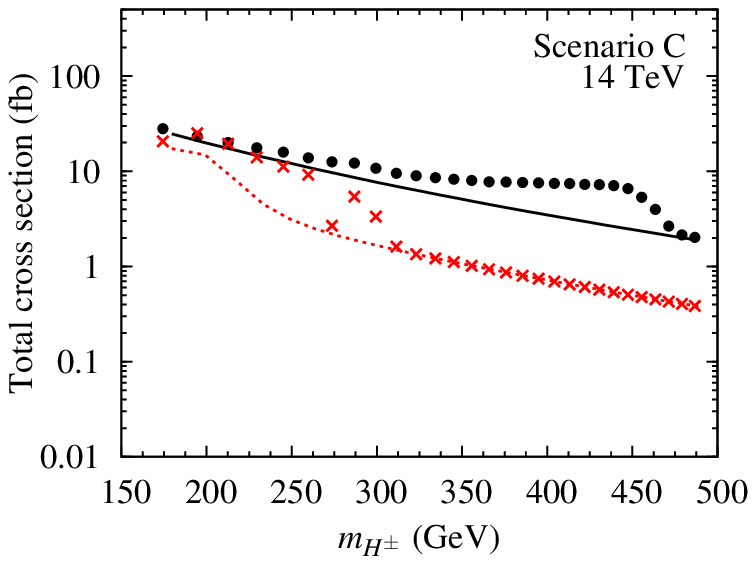}
   \caption{Total hadronic cross section $pp\to H^\pm W^\mp$ at $\sqrt{s}=7$~TeV (left) and $\sqrt{s}=14$~TeV (right) as a function
of the charged Higgs boson mass in scenarios A--C (with variable
$A_\lambda$). The symbols indicate the separate contributions to the
NMSSM cross section from $b\bar{b}$ (dots) and $gg$ (crosses), while
the lines show the corresponding contributions ($b\bar{b}$ solid,
$gg$ dotted) in the MSSM limit.}
 \label{fig:A_had}
\end{figure}
For the numerical evaluation of the cross sections we consider $pp$ collisions at the LHC at the two center of mass energies $\sqrt{s}=7$~TeV and $\sqrt{s}=14$~TeV. We use CTEQ6.6 parton distributions \cite{Nadolsky:2008zw} and a fixed
$\muF=\MHp+m_W$. The results for Scenarios A--C are presented in \reffig{A_had}, shown in parallel for $7$~TeV (left) and 14~TeV (right). We give the contributions of $b\bar{b}$ (big dots) and $gg$ (crosses) separately, together with their respective contributions in the MSSM limit (as solid and dotted lines, respectively). 
By comparing the NMSSM results to the MSSM limit, it can be seen from these plots that it is possible to have substantial NMSSM enhancements also in the hadron-level cross section.
In some mass ranges this enhancement can be an order of magnitude. The observed shapes, with
an almost flat dependence of the cross section on the charged Higgs mass up to some mass,
and then a rapid fall-off to the MSSM value, is caused by the contribution of the resonances.
In the region where  the $H^\pm W^\mp$ threshold is below the resonance mass (cf.~\reffig{hmasses}), the cross section is dominated by the contribution of resonant diagrams. When the transition takes place, the non-resonant behavior quickly becomes similar to the MSSM case. Note that the contributions from $b\bar{b}$ and $gg$ in general receive their largest resonant enhancements from two different resonances, as discussed in Section~\ref{sect:parton}. This also leads to the two different $\MHp$ regions where to cross section is flat for the two cases.
Similarly to the discussion of the partonic cross sections above, we can see how the three scenarios A--C differ mainly in the position of the resonances, which results in  a cross section enhancement (for $b\bar{b}$) at low $\MHp$ in Scenario A, and towards higher masses in Scenario C. The opposite is true for the $gg$ contribution. 

\begin{figure}[t!]
\includegraphics[width=0.45\columnwidth]{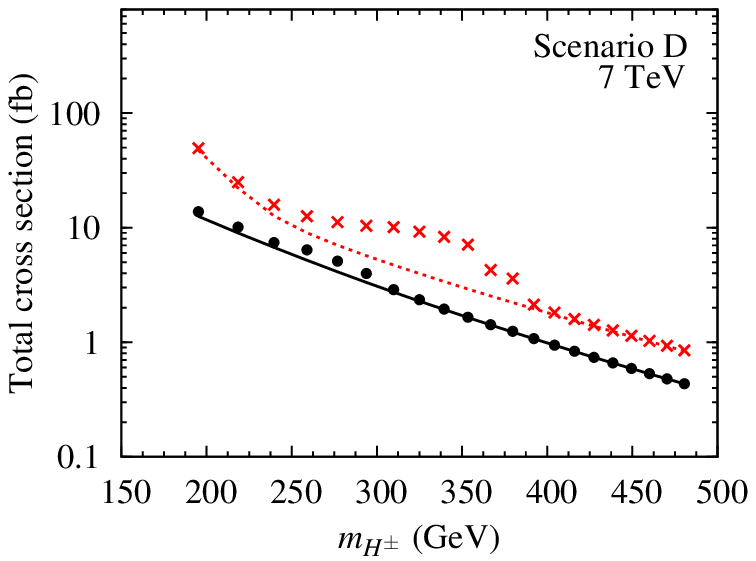}
\includegraphics[width=0.45\columnwidth]{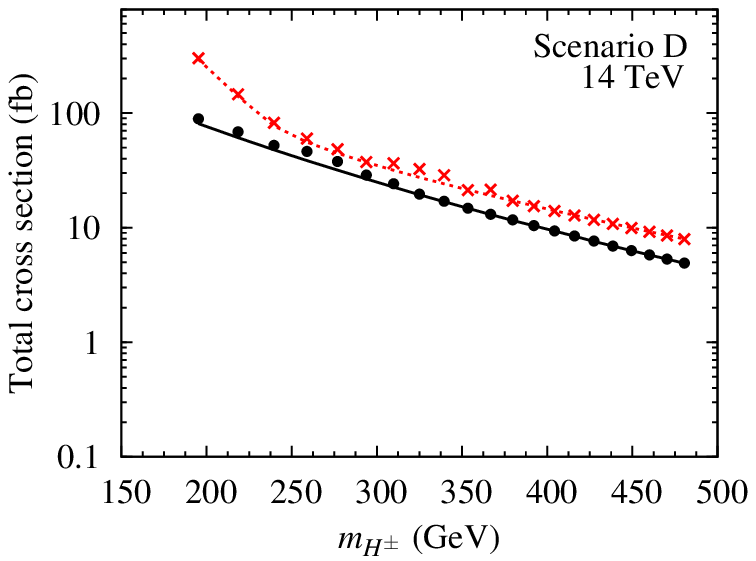}
   \caption{Total hadronic cross section $pp\to H^\pm W^\mp$ at $\sqrt{s}=7$~TeV (left)
 and $\sqrt{s}=14$~TeV (right) for NMSSM Scenario~D. The different
symbols used are explained in the caption of
Figure~\ref{fig:A_had}.}
 \label{fig:D_had}
\end{figure}
\begin{figure}[th!]
\includegraphics[width=0.45\columnwidth]{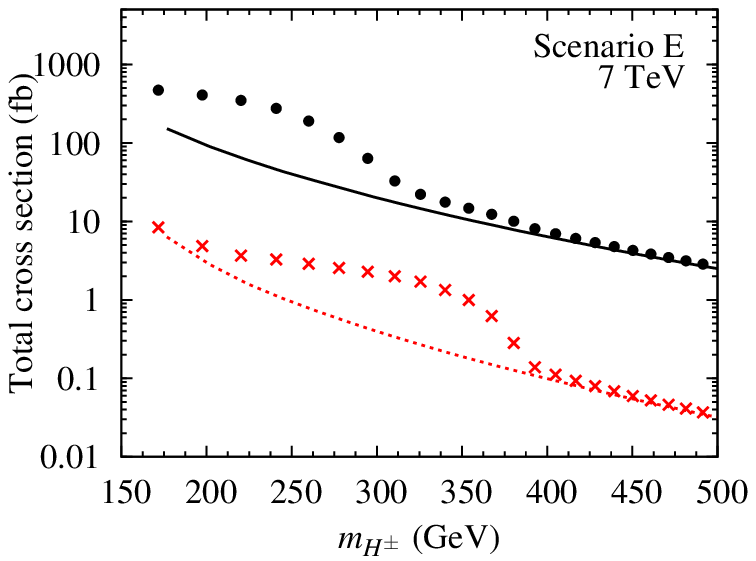}
\includegraphics[width=0.45\columnwidth]{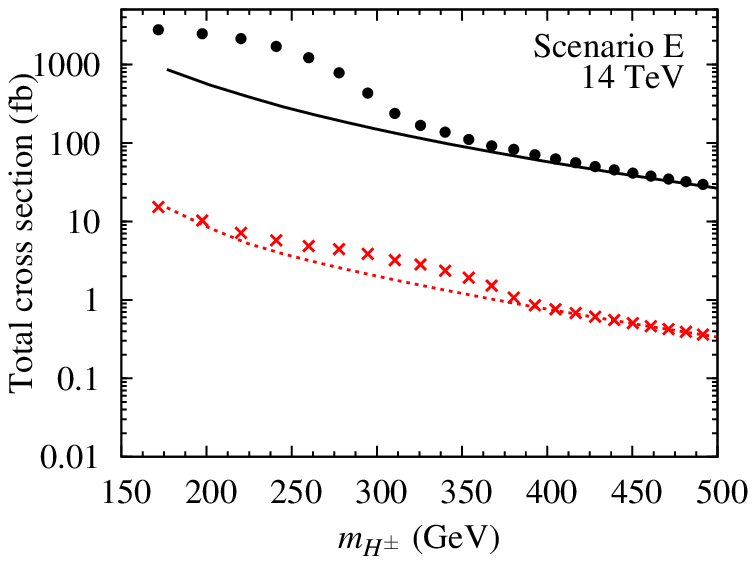}
   \caption{Total hadronic cross section $pp\to H^\pm W^\mp$ at $\sqrt{s}=7$~TeV (left) and $\sqrt{s}=14$~TeV (right) for NMSSM Scenario~E. The different symbols used are explained in the caption of Figure~\ref{fig:A_had}.}
 \label{fig:E_had}
\end{figure}

In \reffig{D_had} we show the hadronic cross section for Scenario D. We see from this figure that the low value of $\tan\beta$ in this scenario gives a significantly larger $gg$ contribution to the total cross section than in the other cases; it dominates over $b\bar{b}$ for the full range of $\MHp$. There is more room for a large $gg$ contribution in the NMSSM, since lower $\tan\beta$ is still allowed in this model compared to the MSSM. We note a relatively small resonance enhancement in this scenario, in particular for $b\bar{b}$ where it is barely visible. 
For our last scenario, Scenario E, the cross section is shown in \reffig{E_had}.
With a large $\tan\beta$, this scenario has been selected to maximize the Higgs couplings to $b\bar{b}$, and indeed we find the largest total cross section for this scenario. As could be expected, the $b\bar{b}$ contribution dominates completely, being almost two orders of magnitude larger than $gg$ for $\MHp>400\GeV$. In this scenario the resonances are again pronounced, and we observe clearly the coupling of $b\bar{b}$ to the (lighter) $A_2$, while the $gg$ contribution is more efficiently enhanced by the (heavier) $H_3$ resonance.

It is interesting to note from the hadron-level results that the relative importance of $b\bar{b}$ and $gg$ may change drastically in regions where either of the resonances dominate. This change in the relative contributions can be much larger than what is expected from the energy scaling of the parton luminosities, cf.~Fig. \ref{fig:lumiratio}. It is a result of the different couplings of the $H_3$ and $A_2$ resonances to the initial and final states. In some of our scenarios the $gg$ contribution can even become larger than the $b\bar b$ contribution in certain regions of $\MHp$ (in Scenario D, the neutral Higgses have a reduced coupling to $b$ quarks, and the $gg$ contribution is dominant for all values of $\MHp$).

Another feature of the $gg$ contribution to the hadronic cross section which deserves a comment is the apparent larger NMSSM enhancement at the 7 TeV LHC compared to the 14 TeV case. The shape of the enhancements is due to a convolution of the resonance peaks observed in the parton-level cross section with the dependence of the PDFs on the partonic center-of-mass energy $\sqrt{\hat s}$ (including the $x_{1,2}$-dependence of the PDFs). There can also be interference effects between the resonant contributions and (non-resonant) boxes.
\begin{figure}
\includegraphics[width=0.45\columnwidth]{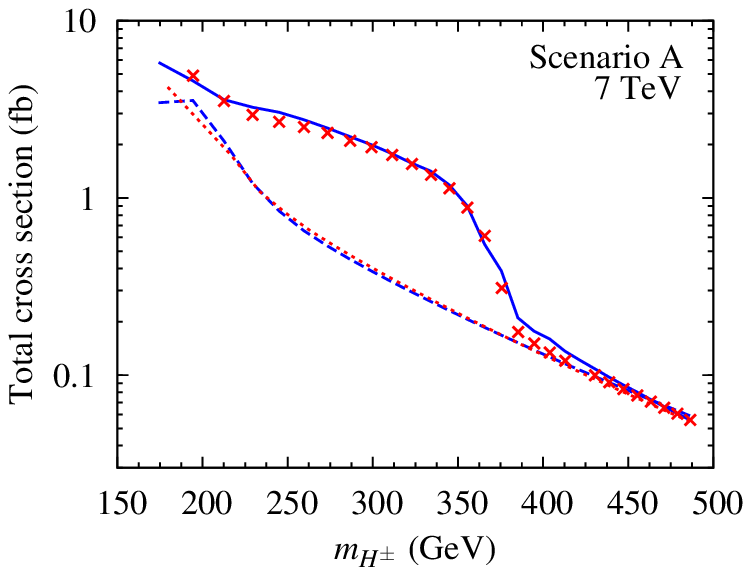}
\includegraphics[width=0.45\columnwidth]{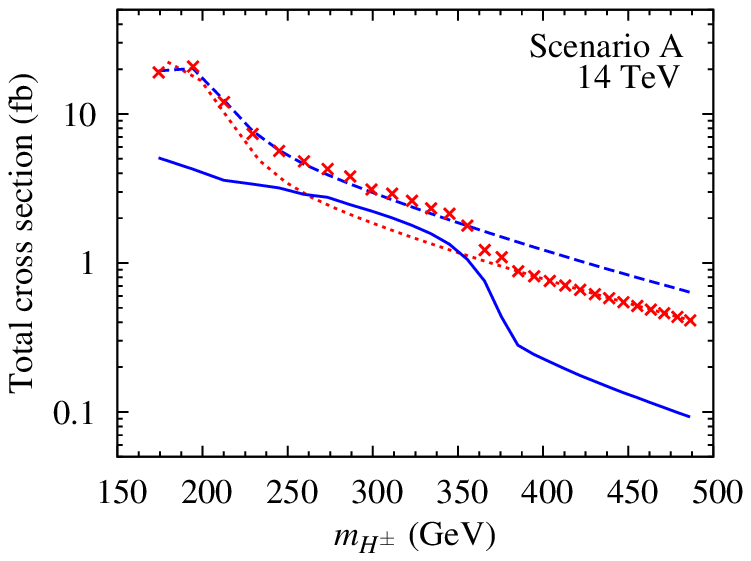}
   \caption{The gluonic contribution to the total hadronic
   cross section $pp \to H^\pm W^\mp$ at
 $\sqrt{s}=7$~TeV (left) and $\sqrt{s}=14$~TeV (right) for NMSSM
Scenario~A (crosses) compared to the MSSM limit (dotted line). 
The blue lines correspond to the separate contributions to $gg\to H^\pm W^\mp$ from only including triangles (solid line) and box diagrams (dashed).}
 \label{fig:had_detail}
\end{figure}
To investigate the relative importance of the different contributions in some more detail, we illustrate in Fig.~\ref{fig:had_detail} the contributions from (boxes)$^2$ and (triangles)$^2$ to the total cross section for Scenario A.
At $\sqrt{s}=7$ TeV, we see that the box and triangle contributions are similar in the non-resonant region, i.e.~at relatively large charged Higgs masses $\MHp\gtrsim 380\GeV$.
Interestingly enough, the full cross section is numerically at the
same level as separately the box and triangle contributions, which means
that the destructive interference between the two is large and
similar to that in MSSM. However, in the resonant region (corresponding to $\MHp< 380\GeV$), the triangle contribution becomes much more pronounced and strongly enhances the total cross
section. The interference effects are naturally quite small there.
At higher energies, 
$\sqrt{s}=14$ TeV, we observe a somewhat different picture in Fig.~\ref{fig:had_detail}. The boxes in this case gives a large (dominant) contribution to the total gluon-initiated cross section over the whole range in $\MHp$. In the non-resonant region, $\MHp\gtrsim 380$ GeV, the interference with the smaller triangle contribution noticeably
decreases the cross section compared to the box contribution alone. In
the resonant region, $\MHp<380$ GeV, the triangles become more
important, but remain sub-dominant compared to the boxes. The cross section is therefore only  enhanced slightly with respect to the MSSM case. 
Analogously to the $7$~TeV case, the interference turns out to be less important than in the non-resonant region.

\section{Collider phenomenology}
We have shown that the total cross sections for
associated $H^\pm W^\mp$ production in the NMSSM can be substantially enhanced
compared to the corresponding cross sections in the MSSM. We would
now like to discuss the phenomenological implications of this for
searches at the LHC. Dedicated collider studies of this channel have
been performed for MSSM and two-Higgs doublet models
\cite{Moretti:1998xq,Eriksson:2006yt,Gao:2007wz,Hashemi:2010ce,Bao:2011sy}.
While we will comment here on what features of these studies that are
relevant for NMSSM---and what features will be different---we
leave a dedicated phenomenological study for the future, since this would require
calculation of differential cross sections and consideration of backgrounds and sensitivities.

The defining feature of the various possible ways of detecting
associated production is how the charged Higgs decays. The previous
studies have all concentrated on the decays that are relevant in the
MSSM or 2HDM, namely $H^+\to\tau^+\nu,c\bar s$ for light charged
Higgs bosons and $H^+\to t\bar b$ for heavy charged Higgs. Decays to SUSY particles
may also be important for heavy $H^\pm$.
As we discussed above, in the NMSSM the decay $H^\pm \to W^{\pm}
A_1$ is sometimes dominant, and will lead to quite different
experimental requirements. This channel is normally not possible in
the MSSM, since $m_A$ and $\MHp$ are close to degenerate. In the
NMSSM, the decay width is proportional to the
doublet component of $A_1$, but may be large even if the $A_1$ is
mostly singlet~\cite{Akeroyd:2007yj,Mahmoudi:2010xp}. The $A_1$
boson can be very light in the NMSSM, and in such a case its
dominant decays will be $A_1\to b\bar b$ or $A_1\to \tau^+\tau^-$. We do not
consider scenarios with a light $A_1$ in this paper, but they could nevertheless be
of interest.

\begin{figure}[t!]
\includegraphics[width=0.32\columnwidth]{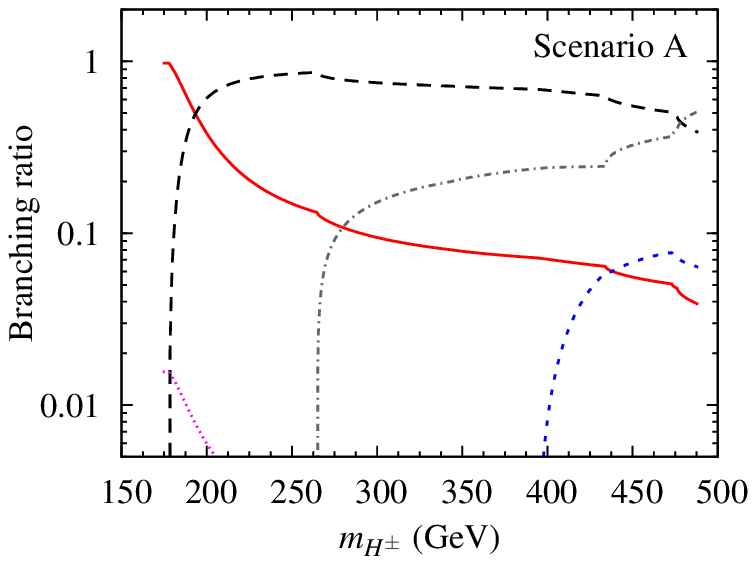}
\includegraphics[width=0.32\columnwidth]{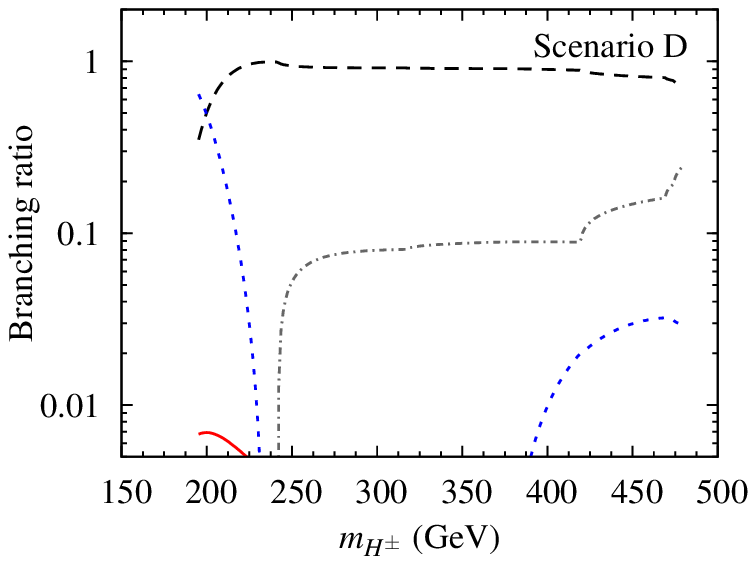}
\includegraphics[width=0.32\columnwidth]{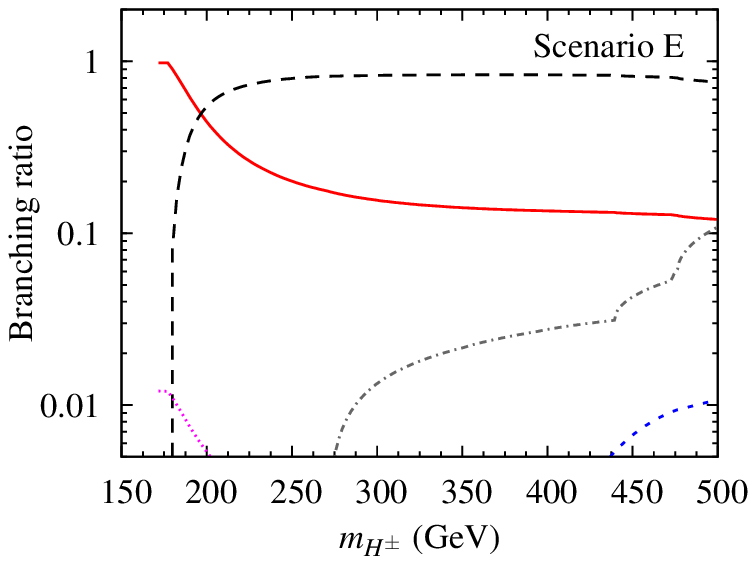}
\caption{Charged Higgs branching ratios $H^+\to \tau^+\nu$ (solid, red),
$H^+\to t\bar{b}$ (long dashes, black), $H^+\to c\bar{b}$ (dotted,
purple), $H^+\to W^+A_1$ (short dashes, blue), and $H^+\to
\mathrm{SUSY}$ (dot-dashed, gray) for NMSSM benchmark scenarios A
(left), D (center), and E (right).} \label{fig:hdecay}
\end{figure}
In Fig.~\ref{fig:hdecay} we show the decay branching ratios of
$H^+$, calculated using \NMT, for three of our benchmark scenarios.
Since the most important parameter entering the determination of the
decay modes is $\tan\beta$, the results for scenarios B and C are
very similar to those for Scenario A and we do not show them
explicitly. All our scenarios have, as explained
above, charged Higgs masses above the threshold for the decay
$H^+\to t\bar b$. As can be seen from
Fig.~\ref{fig:hdecay}, the decay $H^+\to t\bar b$ dominates over a
wide range in masses, but for scenarios A and E the decay $H^+\to
\tau^+\nu$ is appreciable over the entire mass ranges plotted, 
and dominates close to the $t\bar b$ threshold. The $H^\pm\to W^\pm A_1$
decay is only relevant for Scenario D, which has a somewhat lighter
$A_1$, or for very heavy $H^+$ in scenarios A--C. SUSY decays to a
chargino-neutralino pair,
$H^+\to\widetilde\chi_1^+\widetilde\chi_1^0$, also become appreciable
above threshold, but this of course depends very much on the parameters entering the neutralino and chargino sectors. To summarize, the situation for our scenarios is thus similar to the MSSM case in terms of decay channels, except possibly for Scenario D. However,
the possible Higgs masses and the size of the cross sections can be
different. We also would like to emphasize that, for a complete coverage of the phenomenological possibilities in the NMSSM, other scenarios where the decay modes are different should be considered.

Previous collider studies, which were all based on the MSSM, have thus used either the $H^+\to \tau^+\nu$
or the $H^+\to t\bar b$ decay channels. We will now make some simple estimates of how these results translate to the NMSSM. Keep in mind that although we can draw some general conclusions based on these estimates, there may be more subtle differences, caused e.g.~by different dependences of the differential cross sections on kinematical variables when the relative contributions of different types of processes are different.

 In \cite{Eriksson:2006yt}, the $H^+\to \tau^+\nu$ channel was considered
with hadronic $\tau$ and $W^\mp$ decays. The main background to this
signal is the production of $W+2$~jets. A parton-level study showed
that for $\MHp=175\GeV$ in the $m_h^\text{max}$ scenario
at $\tan\beta=50$, it should be possible after a series of cuts to
get a statistically significant signal at the 14 TeV LHC with
300~fb$^{-1}$ of data. In this scenario, the total production cross
section was 55 fb, with a charged Higgs branching ratio into $\tau^+\nu$ of 100\%,
and the obtained $S/\sqrt{B}$ value was 17. We may rescale this
result to e.g.\ our Scenario A (with a much lower $\tan\beta$), where the cross section is,
conservatively, $100$~fb and the branching ratio $\mathrm{BR}(H^+\to\tau^+\nu)=100\%$, which would
yield a significance of about $S/\sqrt{B}=30$. A scenario with
$\MHp=400$~GeV was also studied \cite{Eriksson:2006yt}, but here
the significance was only $S/\sqrt{B}=2.6$ due to the lower cross
section and branching ratio. At this higher mass, we do not predict
an enhancement of the cross section in the NMSSM for Scenario A, but
rather a similar cross section to the one in the MSSM; our expected significance should therefore be similar. If we instead consider Scenario C, which has the resonant enhancement at a higher mass, the cross section at $\MHp=400\GeV$ is $10$~fb, while the branching ratio is about 10\%
(instead of 20\%). This would yield a higher significance, but still
not above $S/\sqrt{B}=5$.
Similar conclusions as those presented in \cite{Eriksson:2006yt} were found in\cite{Gao:2007wz}, and more recently in \cite{Hashemi:2010ce}, where a detailed
study of discovery contours in the MSSM was performed using both
leptonic and hadronic $W$ decays.

The $H^+\to t\bar b$ decay was first studied in \cite{Moretti:1998xq}, where hadronic decays of the top quark and the $W$ from the top quark, and leptonic decay of the other $W$
were considered. The hadronic decay of the $\hp$ has the advantage
over the $\tau\nu$ channel that the mass can be completely
reconstructed, while the charged lepton from the other $W$ provides
a useful trigger. The conclusion of this work was
unfortunately that the $t\bar t$ background completely overwhelms
the signal. However, it was recently
argued  \cite{Bao:2011sy} that with the use of optimized cuts this channel can be
useful to obtain significant results. No numbers for the
significance of the $H^+\to t\bar b$ channel have been given in \cite{Moretti:1998xq}, but we note that in our Scenario A, the total cross section times branching
ratio at $\MHp=200$~GeV is roughly 10 times larger than the one
used in the analysis there, and it is therefore plausible that the
significance could be improved.

The remaining, possibly useful, decay channel is $H^\pm\to W^\pm A_1$,
followed by $A_1\to b\bar b$ or $A_1\to \tau^+\tau^-$ decay. Note that the
final state in the $b\bar b$ case is the same as for the hadronic
$H^+\to t\bar b$ case, but with a $b\bar b$ pair that should
reconstruct the $A_1$ mass. This may provide an additional handle on
the signal, but it is not obvious that this is useful
experimentally; it may be that the $\tau^+\tau^-$ decay proves more
useful. 
If possible, we would like to suggest to the experiments that $t\bar t$ samples with leptonic $W$ decays are investigated for $b\bar b$ resonances. To look for (low mass) $b\bar{b}$ resonances in more energetic events has also been proposed recently as a means of searching for light, singlet-dominated, NMSSM Higgs bosons produced in SUSY cascades~\cite{Stal:2011cz}.
In any case, we think that the $H^\pm\to W^\pm A_1$ channel deserves a detailed study, and as far as we know this has not been performed for the case of $H^\pm W^\mp$ production.

\section{Summary and conclusions}
We have studied associated charged Higgs and $W$ boson production in the NMSSM. This process is complementary to the main production modes anticipated for heavy charged Higgs bosons ($\MHp>m_t$) at the LHC. We calculated the leading order contributions to the total hadronic cross section $pp\to H^\pm W^\mp$ in a general NMSSM setting, corresponding to the tree-level $b\bar{b}$ contribution and the $gg$-initiated subprocess at one loop. The calculation has been performed using an improved Born approximation, taking into account the most important effects of (S)QCD higher order corrections.

For $H^\pm W^\mp$ production in the NMSSM, we have first investigated the parameter dependence of the parton-level cross section and the corresponding result in the MSSM limit. Starting from a maximal mixing scenario, we have then defined five NMSSM benchmark scenarios, with the common feature that they allow for resonant contributions from heavy singlet-like Higgs bosons. These resonances couple both to the $b\bar{b}$ and $gg$ initial states, and we find that they can lead to significant enhancements of the cross section (by up to an order of magnitude) over a wide range of charged Higgs masses in the benchmark scenarios. The presence of these resonances is a genuine feature of the NMSSM, since Higgs mass configurations like this are not possible in the MSSM. This process might therefore be useful as a discriminator between the two models.

We also discussed briefly the phenomenological implications of $H^\pm W^\mp$ production in the NMSSM. Based on previous work for the MSSM, we estimated the discovery significance that could be expected for different decay channels of $\hp$. From these estimates it seems likely that the chances of detecting the charged Higgs boson of the NMSSM at the LHC through this process are quite good, at least in scenarios similar to ours with a not too heavy $\hp$. It may not even be necessary to wait for the 300~fb$^{-1}$ at $14$~TeV that were previously assumed in MSSM studies, but clearly it will still require more data than ``standard'' Higgs searches. However, it should be remembered that these results are rough estimates based on rescaling of earlier MSSM results with the total cross section ratio. More detailed Monte Carlo studies should therefore be done, for example on the impact of the different kinematics between the resonant and non-resonant contributions. It would also be necessary to revise previous results for the present and future running conditions of the LHC, something which is also true for the existing MSSM studies. To more seriously assess the prospects of observing associated $H^\pm W^\mp$ production at the LHC it would eventually be necessary to perform an analysis at the level of full experimental detector simulation. Perhaps the encouragement provided by these positive results could trigger the interest necessary to revisit this channel as a probe for the NMSSM and other models beyond the MSSM.

\acknowledgments \noindent Helpful discussions with Johan Rathsman
are gratefully acknowledged. R.E.~was supported by the Swedish
Research Council under contract 2007-4071. R.P.~was supported by the Carl
Trygger Foundation (Sweden). O.S.~was supported by the
Collaborative Research Center SFB676  of the DFG, ``Particles,
Strings, and the Early Universe''.

\bibliographystyle{apsrev}
\bibliography{HW_nmssm}

\end{document}